# Visualizing Exotic Orbital Texture in the Single-Layer Mott Insulator 1T-TaSe$_2$

Yi Chen[1,2,†], Wei Ruan[1,2,†], Meng Wu[1,2,†], Shujie Tang[3,4,5,6,7,†], Hyejin Ryu[5,8], Hsin-Zon Tsai[1,9], Ryan Lee[1], Salman Kahn[1], Franklin Liou[1], Caihong Jia[1,2,10], Oliver R. Albertini[11], Hongyu Xiong[3,4], Tao Jia[3,4], Zhi Liu[6], Jonathan A. Sobota[3,5], Amy Y. Liu[11], Joel E. Moore[1,2], Zhi-Xun Shen[3,4], Steven G. Louie[1,2], Sung-Kwan Mo[5], and Michael F. Crommie[1,2,12,*]

[1]Department of Physics, University of California, Berkeley, CA 94720, USA

[2]Materials Sciences Division, Lawrence Berkeley National Laboratory, Berkeley, California 94720, USA

[3]Stanford Institute for Materials and Energy Sciences, SLAC National Accelerator Laboratory and Stanford University, Menlo Park, CA 94025, USA

[4]Geballe Laboratory for Advanced Materials, Departments of Physics and Applied Physics, Stanford University, Stanford, CA 94305, USA

[5]Advanced Light Source, Lawrence Berkeley National Laboratory, Berkeley, CA 94720, USA

[6]CAS Center for Excellence in Superconducting Electronics, Shanghai Institute of Microsystem and Information Technology, Chinese Academy of Sciences, Shanghai 200050, China

[7]School of Physical Science and Technology, Shanghai Tech University, Shanghai 200031, China

[8]Center for Spintronics, Korea Institute of Science and Technology, Seoul 02792, South Korea

[9]International Collaborative Laboratory of 2D Materials for Optoelectronic Science & Technology of Ministry of Education, Engineering Technology Research Center for 2D Material Information Function Devices and Systems of Guangdong Province, Shenzhen University, Shenzhen 518060, China

[10]Henan Key Laboratory of Photovoltaic Materials and Laboratory of Low-dimensional Materials Science, School of Physics and Electronics, Henan University, Kaifeng 475004, China

[11]Department of Physics, Georgetown University, Washington, DC 20057, USA

[12]Kavli Energy Nano Sciences Institute at the University of California Berkeley and the Lawrence Berkeley National Laboratory, Berkeley, California 94720, USA

*† These authors contributed equally to this work.*

*\*e-mail: crommie@berkeley.edu*

**Abstract:**

Mott insulating behavior is induced by strong electron correlation and can lead to exotic states of matter such as unconventional superconductivity and quantum spin liquids. Recent advances in van der Waals material synthesis enable the exploration of novel Mott systems in the two-dimensional limit. Here we report characterization of the local electronic properties of single- and few-layer 1T-TaSe$_2$ via spatial- and momentum-resolved spectroscopy involving scanning tunneling microscopy and angle-resolved photoemission. Our combined experimental and theoretical study indicates that electron correlation induces a robust Mott insulator state in single-layer 1T-TaSe$_2$ that is accompanied by novel orbital texture. Inclusion of interlayer coupling weakens the insulating phase in 1T-TaSe$_2$, as seen by strong reduction of its energy gap and quenching of correlation-driven orbital texture in bilayer and trilayer 1T-TaSe$_2$. Our results establish single-layer 1T-TaSe$_2$ as a useful new platform for investigating strong correlation physics in two dimensions.

Two-dimensional (2D) Mott insulators emerge when the Coulomb interaction ($U$) exceeds the bandwidth ($W$) in partially-filled band systems that can be described by 2D Hubbard-like models[1]. Correlated electronic behavior in quasi-2D Mott insulators leads to novel collective quantum phenomena[2,3] such as high-temperature superconductivity which is widely believed to arise through doping of Mott insulating copper-oxygen layers[4,5]. Stacked graphene systems have also recently been found to exhibit Mott-like insulating behavior and unconventional superconductivity upon gating[6-9]. Layered transition metal dichalcogenides (TMDs) offer another family of correlated quasi-2D materials, two examples being bulk 1T-TaS$_2$ and the surface of bulk 1T-TaSe$_2$ which have long been known to host unusual insulating phases in the star-of-David charge density wave (CDW) state[10-13]. Although widely believed to be Mott insulators[11,14,15], the insulating nature of these bulk systems is complicated by interlayer CDW stacking whose effects on the insulating phase remain controversial[16]. Interlayer hopping (which increases $W$) and interlayer dielectric screening (which decreases $U$) are expected to suppress Mott insulating behavior[1,17,18], but orbital stacking has also been predicted to open a hybridization gap even in the absence of electron correlation[16,19,20].

Atomically-thin 1T-TMDs offer an ideal platform to differentiate the contributions of electron correlation and interlayer effects in quasi-2D materials since correlation effects in single-layer systems can be fully characterized in the absence of interlayer coupling. The effect of interlayer coupling can then be systematically determined by adding new layers to a material one at a time and mapping out the resulting stacking order and wavefunction properties. Previous studies on single-layer 1T-NbSe$_2$ and 1T-TaSe$_2$ have found unusual insulating behavior[21,22], but electronic wavefunction texture and interlayer coupling effects were not examined. The nature of the insulating phase in these single-layer materials thus remains inconclusive.

Here we report a combined scanning tunneling spectroscopy (STS), angle-resolved photoemission spectroscopy (ARPES), and theoretical study of the electronic structure of single-layer (SL) 1T-TaSe$_2$. Our results show that in the absence of interlayer coupling SL 1T-TaSe$_2$ hosts a Mott-insulating ground state that exhibits a 109 ± 18 meV energy gap and exotic orbital texture. Bilayer and trilayer 1T-TaSe$_2$ with shifted stacking order exhibit successively smaller energy gaps and show no signs of the novel orbital texture seen in the SL limit. The SL band structure and density of states (DOS) of 1T-TaSe$_2$ are found to be consistent with DFT+U calculations, confirming its Mott insulator nature. The exotic SL orbital texture, however, is not captured by DFT+U, but is consistent with the behavior expected for a weakly screened, strongly correlated 2D insulator. Reduction of the 1T-TaSe$_2$ bandgap and quenching of the novel orbital texture by the addition of new layers shows that the effect of interlayer coupling on shifted-stacked 1T-TaSe$_2$ is to weaken the Mott behavior. The SL limit of 1T-TaSe$_2$ is thus unique in that strong correlation effects here are most pronounced, affecting both the energy gap and electron wavefunction symmetry.

**Electronic structure of single-layer 1T-TaSe$_2$ in the CDW phase**

Our experiments were carried out on 1T-TaSe$_2$ thin films grown by molecular beam epitaxy (MBE) on epitaxial bilayer-graphene-terminated (BLG) 6H-SiC(0001), as sketched in Fig. 1a. The crystal structure of 1T-TaSe$_2$ consists of a layer of Ta atoms sandwiched between two layers of Se atoms in an octahedral coordination. Fig. 1b illustrates the hexagonal morphology of our 1T-TaSe$_2$ islands, indicating high epitaxial growth quality. A triangular CDW superlattice is observed on single-layer, bilayer, and trilayer 1T-TaSe$_2$, as seen in Figs. 1b, c, and Supplementary Fig. 1 where each bright spot corresponds to a star-of-David CDW supercell. Fourier analysis of STM images (Supplementary Fig. 2) together with low-energy electron diffraction (LEED) patterns (Supplementary Fig. 3) confirm the formation of a $\sqrt{13} \times \sqrt{13}$ CDW in SL 1T-TaSe$_2$, similar to the commensurate CDW phase of bulk 1T-TaSe$_2$ at $T < 473$ K[23] (the atomic lattice and CDW superlattice are observed to have a relative rotation angle of 13.9°). Reflection high-energy electron diffraction (RHEED) patterns (Fig. 1e) and X-ray photoelectron spectroscopy (XPS) (Fig. 1f) show the structural and chemical integrity of our single-layer 1T-TaSe$_2$ samples (1T and 1H islands do coexist in our samples due to the metastability of 1T-TaSe$_2$ (Supplementary Fig. 4)).

We experimentally determined the electronic structure of SL 1T-TaSe$_2$ in the star-of-David CDW phase using ARPES and STS. Figs. 2a and 2b show the ARPES spectra of SL 1T-TaSe$_2$ for *p*- and *s*-polarized light, respectively, obtained at $T = 12$ K. At low binding energies (-0.15 eV $< E - E_F < 0$) the SL 1T-TaSe$_2$ ARPES spectra show strongly diminished intensity at all observed momenta, indicating insulating behavior (some ARPES intensity from coexisting 1H-TaSe$_2$ islands can be seen crossing $E_F$ at $k \approx 0.5$ Å$^{-1}$ [24] (white dashed lines)). The CDW superlattice potential induces band folding into a smaller CDW Brillouin zone (BZ) (Fig. 2b inset). One such band can be seen in the ARPES spectrum for *p*-polarized light (Fig. 2a) which shows a prominent flat band centered at $E - E_F \approx -0.26$ eV within the first CDW BZ (black dashed box). A more dispersive band can be resolved outside of the first CDW BZ boundary (vertical dashed lines labeled A and B mark this boundary). These features have no obvious photon-energy dependence (Supplementary Fig. 5) and are similar to bands observed by ARPES at the surface of bulk samples of 1T-TaS$_2$[16,25] and 1T-TaSe$_2$[12]. For *s*-polarized light (Fig. 2b) the flat band is much less visible and a manifold of highly dispersive bands near the Γ-point dominates the spectrum.

Our STM d$I$/d$V$ spectrum (Fig. 3a (black curve)) confirms the insulating nature of SL 1T-TaSe$_2$. The electronic local density of states (LDOS) reflected in d$I$/d$V$ exhibits a pronounced valence band peak at $V$ = -0.33 V (labeled V$_1$) while dropping steeply at higher energy until reaching zero at $V \approx$ -0.05 V. The zero LDOS region bracketing the Fermi level yields an energy gap of magnitude 109 ± 18 meV (see Supplementary Fig. 6 for gap determination). The experimental LDOS does not rise again until a narrow conduction band peak is observed centered at $V$ = 0.20 V (labeled C$_1$) in the empty state regime. The LDOS drops to zero again above the C$_1$ peak until higher-lying conduction band features are seen to rise at $V$ > 0.45 V (e.g., C$_2$, C$_3$). Aside from spatial variation in the relative peak heights, this gapped electronic structure is observed uniformly over the entire SL 1T-TaSe$_2$ surface (Supplementary Fig. 7).

**Experimental orbital texture of single-layer 1T-TaSe$_2$**

To gain additional insight into the insulating ground state of SL 1T-TaSe$_2$, we performed d$I$/d$V$ spatial mapping of its energy-dependent orbital texture at constant tip-sample separation (Figs. 3b-h). d$I$/d$V$ maps measured at negative biases all display a similar pattern where high-intensity LDOS is concentrated near the center of each star-of-David (Figs. 3b, c, and Supplementary Fig. 8). The experimental empty-state LDOS of SL 1T-TaSe$_2$, however, exhibits a completely different orbital texture. This is most dramatically seen in the d$I$/d$V$ map taken at the lowest conduction band peak C$_1$ (Fig. 3d) where the center of each CDW supercell is observed to be dark (i.e., no LDOS intensity). At this energy the LDOS exhibits a novel, interlocked "flower" pattern (circled by yellow dashed lines) consisting of six well-defined "petals" (i.e., bright spots) located around the outer rim of each star-of-David. This appearance is completely different from previous reports of conduction band LDOS in bulk 1T-TaSe$_2$ and 1T-TaS$_2$[13,26] (which show LDOS concentrated in the star-of-David centers), and is clearly not due to defects since it follows the CDW periodicity. The 6-fold petal structure has a different symmetry than the 3-fold arrangement of top-layer Se atoms in the spaces between each star-of-David, but it shares the 6-fold symmetry of the Ta atom arrangement (Supplementary Fig. 9).

The d$I$/d$V$ map obtained at a slightly higher bias of $V$ = 0.6 V (C$_2$) show LDOS that is related to the flower pattern in that it exhibits a nearly *inverse* flower (i.e., dark areas at C$_1$ are bright at C$_2$, see circled regions in Fig. 3e). At even higher energies the SL 1T-TaSe$_2$ LDOS displays other intricate orbital textures. The map at 0.8 V (Fig. 3f), for example, shows quasi-triangular patterns with intensity distributed near the outermost Ta C-atoms (labeled according to the convention shown in Fig. 1d). This evolves into trimer-like features at 1.1 V (Fig. 3g), and a network of "rings" with intensity near Ta B-atoms at $V$ = 1.2 V (Fig. 3h) (a complete set of constant-height d$I$/d$V$ maps is shown in Supplementary Fig. 8).

To help quantify the complex energy-dependent LDOS distribution of 1T-TaSe$_2$, we cross-correlated our d$I$/d$V$ maps with the reference map taken near the maximum of the valence band peak. For SL 1T-TaSe$_2$ the reference was taken at $V$ = -0.33V (V$_1$), which exhibits LDOS dominated by inner Ta A- and B-atoms. The resulting cross-correlation values are color-coded in Fig. 3a and show that occupied states (-1V < $V$ < 0 V) all have a strong, positive cross-correlation (blue) with the valence band map at V$_1$ (i.e., the central Ta A- and B-atoms are bright at these energies and the C-atoms are darker). The empty-state cross-correlation, however, is very different. At C$_1$ (where the flower pattern is observed) the LDOS map is strongly anti-correlated (red) with the valence band map since the LDOS here is dominated by Ta C-atoms. At slightly higher energy (C$_2$) the cross-correlation flips to blue. This is due to the LDOS inversion that occurs at this energy (i.e., the inverse flower pattern) which creates intensity at the interior

A- and B-atoms. At higher energy the cross-correlation flips again to red and stays red over a fairly wide energy range (~0.5 eV) before flipping again to blue at $C_3$.

**Energy gap reduction and quenching of exotic orbital texture in few-layer 1T-TaSe$_2$**

We examined the effect of interlayer coupling on 1T-TaSe$_2$ by studying the evolution in electronic structure as 1T-TaSe$_2$ is stacked layer by layer. We first determined the star-of-David CDW stacking order for bilayer and trilayer 1T-TaSe$_2$. As seen in the STM images of Fig. 1b and Supplementary Fig. 1, the CDW stacking order follows the shifted triclinic structure whereby inner Ta "A-atoms" sit on top of outer Ta "C-atoms", similar to stacking observed in bulk 1T-TaSe$_2$[27]. We observe that the energy gap for 1T-TaSe$_2$ narrows dramatically when interlayer coupling is added, as seen in the STM d$I$/d$V$ spectra for bilayer and trilayer 1T-TaSe$_2$ shown in Fig. 4a. The bilayer energy gap reduces to 21 ± 8 meV while trilayer 1T-TaSe$_2$ shows a reduction in LDOS at $E_F$ that can be described as "semimetallic" but exhibits no true energy gap.

In addition to reducing the 1T-TaSe$_2$ energy gap, bilayer and trilayer formation also quenches the exotic orbital texture observed in the single-layer limit. As shown in the insets to Figs. 4b, c, d$I$/d$V$ maps of the lowest conduction band in bilayer and trilayer 1T-TaSe$_2$ show LDOS intensity concentrated near the center of each star-of-David, in stark contrast to the flower-like orbital texture observed in SL 1T-TaSe$_2$ at $C_1$. This difference can also be seen in the color-coded cross-correlation values of bilayer and trilayer 1T-TaSe$_2$ (Figs. 4b, c). Using the valence band LDOS shown in the insets as a reference (which is similar to the SL valence band LDOS of Fig. 3c), the bilayer and trilayer cross-correlation remain strongly positive (blue) throughout the lowest conduction band (thus emphasizing that the LDOS here is concentrated on the interior Ta A- and B-atoms). The distinctive flower pattern seen in SL 1T-TaSe$_2$ at $C_1$ (Fig. 3d) is never seen in bilayer or trilayer LDOS at any bias (Supplementary Figs. 10, 11).

**Theoretical electronic structure of SL 1T-TaSe$_2$ via DFT+U simulations**

There are two main physical questions that we seek to answer regarding our measurements of single- and few-layer 1T-TaSe$_2$. First, what type of insulator is SL 1T-TaSe$_2$? And, second, what is the effect of interlayer coupling on 1T-TaSe$_2$ electronic behavior as new layers are added? To address these questions we first performed a conventional band structure calculation of 1T-TaSe$_2$ using density functional theory (DFT). From an intuitive perspective, SL 1T-TaSe$_2$ is expected to have metallic band structure since there are an odd number of Ta ions in the star-of-David unit cell (13) and each Ta$^{4+}$ ion has only one $d$-electron (charge transfer effects from the graphene substrate are negligible (Supplementary Fig. 12 and Note 1)). As expected, the DFT band structure of SL 1T-TaSe$_2$ in the CDW phase calculated using the GGA-PBE exchange correlation functional shows a metallic half-filled band at $E_F$ (Supplementary Fig. 13). This theoretical result, however, strongly disagrees with our experimental data which shows insulating behavior for SL 1T-TaSe$_2$ (Figs. 2, 3). An explanation for this significant discrepancy is that since the metallic band is so narrow (only ~20 meV wide) it is unstable to splitting into lower and upper Hubbard bands due to a high on-site Coulomb energy ($U$) (i.e., the condition that causes Mott insulators to arise from otherwise metallic phases)[1].

To test for Mott insulator formation in SL 1T-TaSe$_2$ we modeled the effects of electron correlation by performing DFT+U simulations. We find that the DFT+U band structure for a ferromagnetic ground state with $U = 2$ eV reproduces most of our experimentally observed electronic structure for SL 1T-TaSe$_2$ (the DFT+U results were not sensitive to the magnetic ground state and our $U$ value is consistent with previous simulations of related systems[26,28,29] (see Supplementary Figs. 14-16 and Note 2)). The DFT+U band structure was first compared to our ARPES data by unfolding it onto the BZ of an undistorted unit cell. As seen in Fig. 2c it reproduces the gapped electronic structure and shows good overall agreement with the ARPES spectra. In particular, DFT+U predicts that the lower Hubbard band at -0.2 eV originates from Ta $d_{z^2}$ orbitals, consistent with the higher ARPES intensity under *p*-polarized light (Fig. 2a)[30].

The DFT+U simulations were also consistent with much of our STS data as shown in Fig. 5 which displays the simulated DOS spectrum and LDOS maps for SL 1T-TaSe$_2$. The theoretical DOS (Fig. 5a (black line)) reproduces the d$I$/d$V$ spectrum (Fig. 3a) reasonably well in both the occupied and empty states. A lower Hubbard band corresponding to the experimental V$_1$ feature is seen, as well as an upper Hubbard band corresponding to C$_1$, along with higher energy features that correspond to the experimental C$_2$, C$_3$ features. The orbital texture generated by the DFT+U calculations (Figs. 5b-h) also agree well with the experimental d$I$/d$V$ maps in the valence band and upper conduction band regimes (i.e., the energies corresponding to filled states and levels above C$_2$).

However, there are significant discrepancies between the experimental and theoretical LDOS maps in the low-energy conduction band region ($0 < E \lesssim 0.6$ eV) where exotic orbital texture is observed experimentally. This is most dramatically seen by comparing the theoretical upper Hubbard band LDOS map at 0.2 eV (Fig. 5d) with the experimental d$I$/d$V$ map at $V = 0.2$ V (Fig. 3d) (i.e., the energy corresponding to C$_1$). The calculated LDOS has high intensity in the central Ta A- and B-atom regions (similar to what is seen in the lower Hubbard band) while the experiment shows the flower pattern (i.e., the experimental LDOS occupies the Ta C-atom region and is dark in the central area). There also exists significant disagreement at the next higher energy band feature (C$_2$), as seen by comparison of Figs. 3e and 5e. Here the theoretical orbital texture (Fig. 5e) shows propeller-like structures with no central LDOS, while the experimental d$I$/d$V$ map (Fig. 3e) shows an inverse of the C$_1$ flower pattern which has LDOS in the interior region of the star-of-David (a complete set of theoretical LDOS maps is shown in Supplementary Fig. 17).

**Exotic empty-state orbital texture at $C_1$ and $C_2$**

The good agreement between our DFT+U simulations and the majority of our ARPES and STM/STS data provides strong evidence that SL 1T-TaSe$_2$ is a 2D Mott insulator. However, the failure of the simulations to reproduce the exotic conduction band orbital texture at $C_1/C_2$ implies that additional electron-electron interactions occur in SL 1T-TaSe$_2$ that are not captured by DFT+U. Electrons injected from the STM tip into SL 1T-TaSe$_2$ at $C_1/C_2$ experience additional correlation effects originating from their interaction with electrons already present in the occupied electron states. This interaction can be estimated phenomenologically by experimentally mapping the STM tunnel current at a bias value that integrates SL 1T-TaSe$_2$ filled-states, thus allowing their contributions to be converted into an electrostatic potential (Supplementary Note 3). The potential landscape that results from this treatment is repulsive to electrons at the interior of each star-of-David and has minima at precisely the locations of the six-fold $C_1$ flower pattern (Supplementary Fig. 18). This allows the novel orbital texture at $C_1/C_2$ to be understood as a redistribution of the electronic spectral density in the star-of-David center up to higher energy as a result of enhanced Coulomb interactions arising from reduced screening in 2D (i.e., the spectral density that would have been at $C_1$ is pushed up in energy to $C_2$ by electron-electron interactions).

The effect of interlayer coupling on the shifted-stacked 1T-TaSe$_2$ electronic structure is to weaken the Mott insulator phase, both in view of the observed energy gap reduction with increased layer number as well as its effect on orbital texture. The bilayer and trilayer orbital textures, for example, show no signs of the correlation-induced spectral density shift seen in the single-layer material at $C_1/C_2$. Such weakening of the Mott behavior likely arises from an increase in the effective inter-star-of-David hopping parameter ($t$) of the bilayer and trilayer due to interlayer coupling, as well as a reduction in the effective Hubbard $U$ from increased electronic delocalization and screening.

**Outlook**

We have shown that SL 1T-TaSe$_2$ is a strongly correlated 2D Mott insulator characterized by exotic orbital texture. The behavior of this material as observed by ARPES and STM can largely be understood through DFT+U calculations. Novel orbital texture in the upper Hubbard band, however, arises from correlated electron behavior that is not captured by DFT+U and which exists only in the single-layer limit. Interlayer coupling weakens the Mott behavior, thus demonstrating how 1T-TaSe$_2$ evolves into a metal as its thickness is increased layer-by-layer. The robust Mott insulator phase seen in the single-layer limit paves the way for future exploration of exotic quantum phases via new van der Waals heterostructure design and carrier density modification (using both chemical and electrostatic techniques).

**Acknowledgments**

We thank Patrick Lee and Dung-Hai Lee for helpful discussions. This research was supported by the VdW Heterostructure program (KCWF16) (STS measurements and DFT simulations) and the Advanced Light Source (sample growth and ARPES) funded by the Director, Office of Science, Office of Basic Energy Sciences, Materials Sciences and Engineering Division, of the US Department of Energy under Contract No. DE-AC02-05CH11231. Support was also provided by National Science Foundation award EFMA-1542741 (surface treatment and topographic characterization), award DMR-1508412 (development of theory formalism), award DMR-1507141 (electrostatic analysis), and award EFRI-1433307 (CDW model development). The work at the Stanford Institute for Materials and Energy Sciences and Stanford University (ARPES measurements) was supported by the DOE Office of Basic Energy Sciences, Division of Material Science. LEED measurements were supported by the National Natural Science Foundation of China (grant no. 11227902). S. T. acknowledges the support by CPSF-CAS Joint Foundation for Excellent Postdoctoral Fellows. H. R. acknowledges fellowship support from NRF, Korea through Max Planck Korea/POSTECH Research Initiatives No. 2016K1A4A4A01922028. H.-Z.T. acknowledges fellowship support from the Shenzhen Peacock Plan (Grant No. 827-000113, KQJSCX20170727100802505, KQTD2016053112042971).

**Author contributions**

Y.C., W.R., and M.F.C. initiated and conceived the research. Y.C., W.R., H.-Z.T., R.L., S.K., F.L., and C.J. carried out STM/STS measurements and analyses. M.F.C supervised STM/STS measurements and analyses. S.T., H.R., H.X., and T.J. performed sample growth and ARPES measurements. S.-K.M., Z.-X.S., J.A.S., and Z.L. supervised sample growth and ARPES measurements. M.W. performed DFT calculations and theoretical analyses. S.G.L. supervised DFT calculations and theoretical analyses. J.E.M. performed electrostatic modeling. O.R.A. and A.L. provided support for development of the CDW model. Y.C., W.R., and M.F.C. wrote the manuscript with the help from all authors. All authors contributed to the scientific discussion.

**Competing interests**

The authors declare no competing interests.

**Data availability**

The data that support the findings of this study are available from the corresponding author upon reasonable request.

**Methods**

**Sample growth and ARPES measurements**

Single-layer 1T-TaSe$_2$ films were grown on epitaxial bilayer graphene terminated 6H-SiC(0001) substrates in an MBE chamber operating at ultrahigh vacuum (UHV, base pressure 2×10$^{-10}$ Torr) at the HERS endstation of Beamline 10.0.1, Advanced Light Source, Lawrence Berkeley National Laboratory. High purity Ta (99.9%) and Se (99.999%) were evaporated from an electron-beam evaporator and a standard Knudsen cell, respectively, with a Ta:Se flux ratio set between 1:10 and 1:20 and a substrate temperature of 660 °C. A higher substrate temperature (compared with our previous 1H-TaSe$_2$ growth at 550 °C[24]) was used to facilitate the growth of the metastable 1T phase of TaSe$_2$. The growth process was monitored by RHEED. After growth, the films were transferred *in-situ* into the analysis chamber (base pressure 3×10$^{-11}$ Torr) for ARPES and core-level spectra measurements. The ARPES system was equipped with a Scienta R4000 electron analyzer. The photon energy was set at 51 eV (unless specified otherwise) with energy and angular resolution of 12 meV and 0.1°, respectively. *p*- and *s*-polarized light were used, as described elsewhere (ref.[31]). Before taking the films out of vacuum for STM/STS measurements, Se capping layers with ~10 nm thickness were deposited onto the samples for protection. These were later removed by UHV annealing at ~200 °C for 3 hours.

**STM/STS measurements**

STM/STS measurements were performed using a commercial CreaTec LT-STM/AFM system at $T$ = 5 K under UHV conditions. To avoid tip artifacts, STM tips were calibrated on a Au(111) surface by measuring its herringbone surface reconstruction and Shockley surface state before all STM/STS measurements. Both W and Pt-Ir STM tips were used and yielded similar results. STS d$I$/d$V$ spectra were obtained using standard lock-in techniques with a small bias modulation at 401 Hz. The constant-height mode (i.e., feedback loop open) was used for collecting all d$I$/d$V$ conductance maps. Before obtaining each set of maps the STM tip was parked near the sample surface for at least 8 hours to minimize piezoelectric drift effects.

**Electronic structure calculations**

First-principles calculations of single-layer 1T-TaSe$_2$ were performed using density functional theory (DFT) as implemented in the Quantum ESPRESSO package[32]. The onsite Hubbard interaction was added through the simplified rotationally invariant approach using the same $U$ value for each Ta atom[33,34]. A slab model with 16 Å vacuum layer was adopted to avoid interactions between periodic images. We employed optimized norm-conserving Vanderbilt pseudopotentials (ONCVPSP) including Ta 5$s$ and 5$p$ semicore states (with a plane-wave energy cutoff of 90 Ry)[35-37] as well as the Perdew-Burke-Ernzerhof exchange-correlation functional[38] in the generalized gradient approximation (GGA). The structure was fully relaxed until the force on each atom was less than 0.02 eV/Å. The resulting relaxed single-layer 1T-TaSe$_2$ in the $\sqrt{13}\times\sqrt{13}$ CDW phase has a lattice constant of $a$ = 12.63 Å. Spin-orbit coupling was not taken into account in our calculations since it has a negligible influence on the band structure given the inversion symmetry of this system. The unfolding of the band structure from the CDW supercell to the undistorted unit cell was calculated using the BandUP code[39,40].

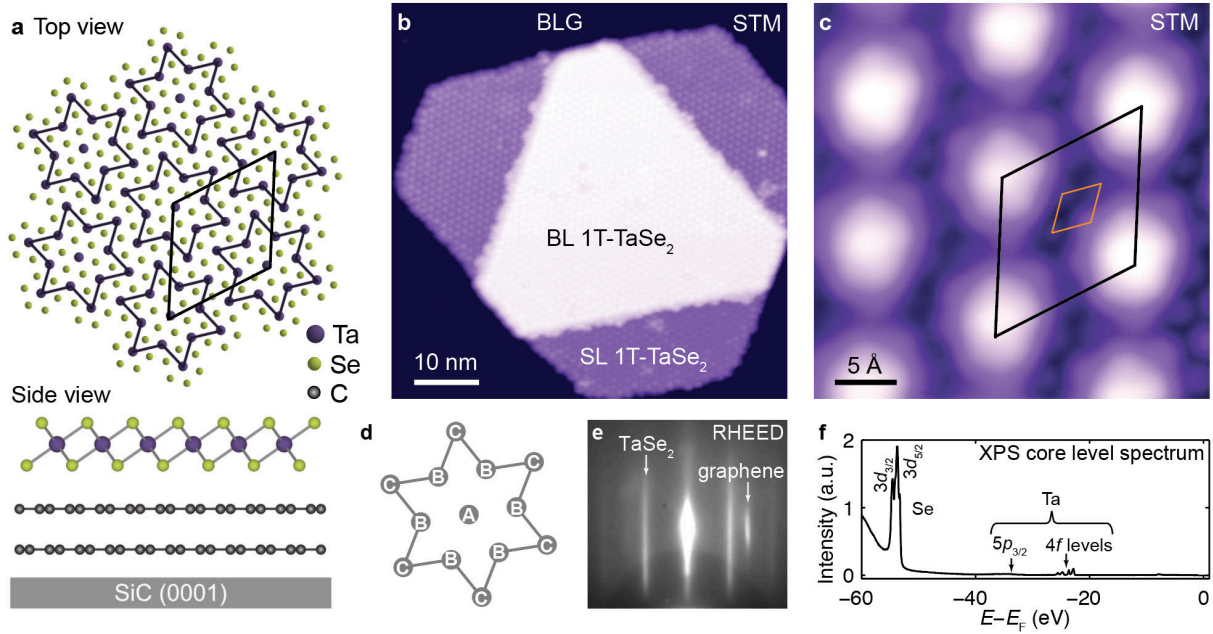

**Fig. 1. Structure of single-layer 1T-TaSe$_2$ in the star-of-David CDW phase. a**, Top and side view sketches of single-layer 1T-TaSe$_2$, including substrate. Clusters of 13 Ta atoms in star-of-David CDW supercells are outlined, as well as the CDW unit cell. **b**, Large-scale STM topograph of a typical 1T-TaSe$_2$ island showing monolayer and bilayer regions ($V_b$ = -0.5 V, $I_t$ = 10 pA, $T$ = 5 K). **c**, A close-up STM image of single-layer 1T-TaSe$_2$. Each bright spot corresponds to a star-of-David supercell ($V_b$ = -0.17 V, $I_t$ = 3 nA, $T$ = 5 K). Black and orange parallelograms mark CDW and atomic unit cells, respectively. **d**, Labels for Ta atoms in the star-of-David CDW supercell depend on radial distance from center. **e**, RHEED pattern of a submonolayer 1T-TaSe$_2$ film. **f**, XPS spectrum showing characteristic peaks of Ta and Se core levels for a submonolayer 1T-TaSe$_2$ film.

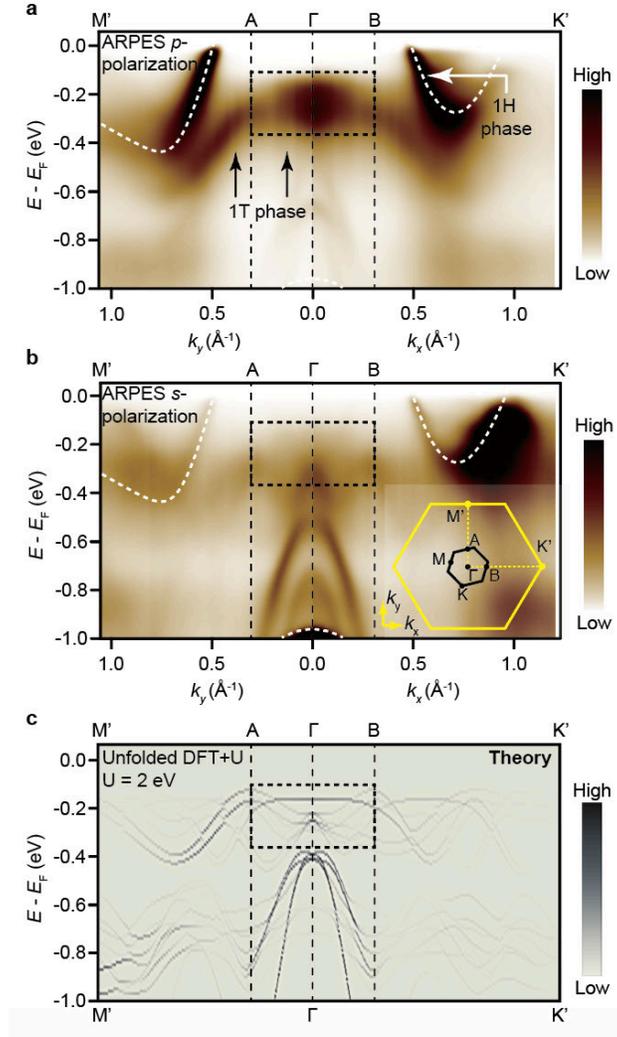

**Fig. 2. ARPES and DFT+U band structure of single-layer 1T-TaSe$_2$.** ARPES spectra acquired with **a,** $p$- and **b,** $s$-polarized light at $T = 12$ K along the Γ-K' and Γ-M' directions defined in the undistorted (i.e., no CDW) unit cell Brillouin zone (yellow hexagon in Fig. 2b inset). ARPES spectra have little intensity at low binding energies except for coexisting 1H-TaSe$_2$ bands that cross $E_F$ at $k \approx 0.5$ Å$^{-1}$ (white dashed lines). A strong flat band is seen under $p$-polarized light in the first CDW Brillouin zone (black dashed box in **a**). The full CDW BZ is sketched in the inset of **b** (black hexagon). **c**, DFT+U band structure ($U = 2$ eV) of single-layer 1T-TaSe$_2$ unfolded onto the undistorted unit cell Brillouin zone shows good agreement with features observed in the ARPES spectra.

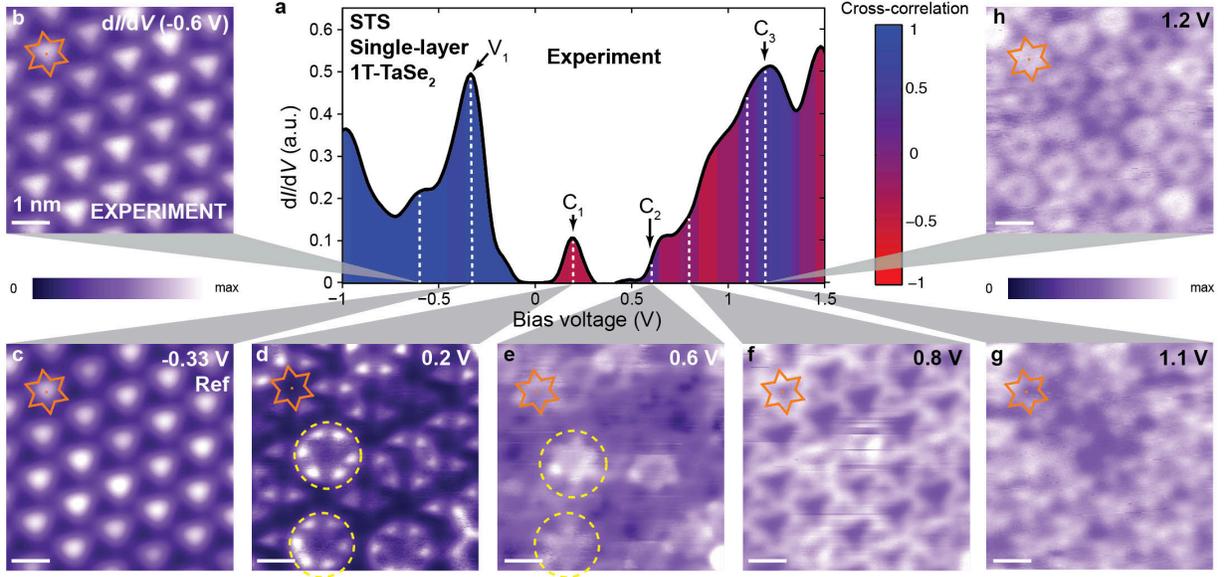

**Fig. 3. Experimental energy-resolved exotic orbital texture of single-layer 1T-TaSe$_2$. a**, STS d$I$/d$V$ spectrum of single-layer 1T-TaSe$_2$ shows a full energy gap bracketed by two STS peaks labeled V$_1$ and C$_1$ ($f$ = 401 Hz, $I_t$ = 50 pA, $V_{RMS}$ = 20 mV). Color shows cross-correlation of d$I$/d$V$ maps at different energies with the reference map shown in **c**. **b-h**, Constant-height d$I$/d$V$ conductance maps of the same area for different bias voltages show energy-dependent orbital texture ($f$ = 401 Hz, $V_{RMS}$ = 20 mV). The same star-of-David CDW supercell is outlined in each map (orange line). Yellow dashed circles in **d**, **e** highlight the exotic LDOS pattern at C$_1$ and C$_2$ and their relative spatial inversion.

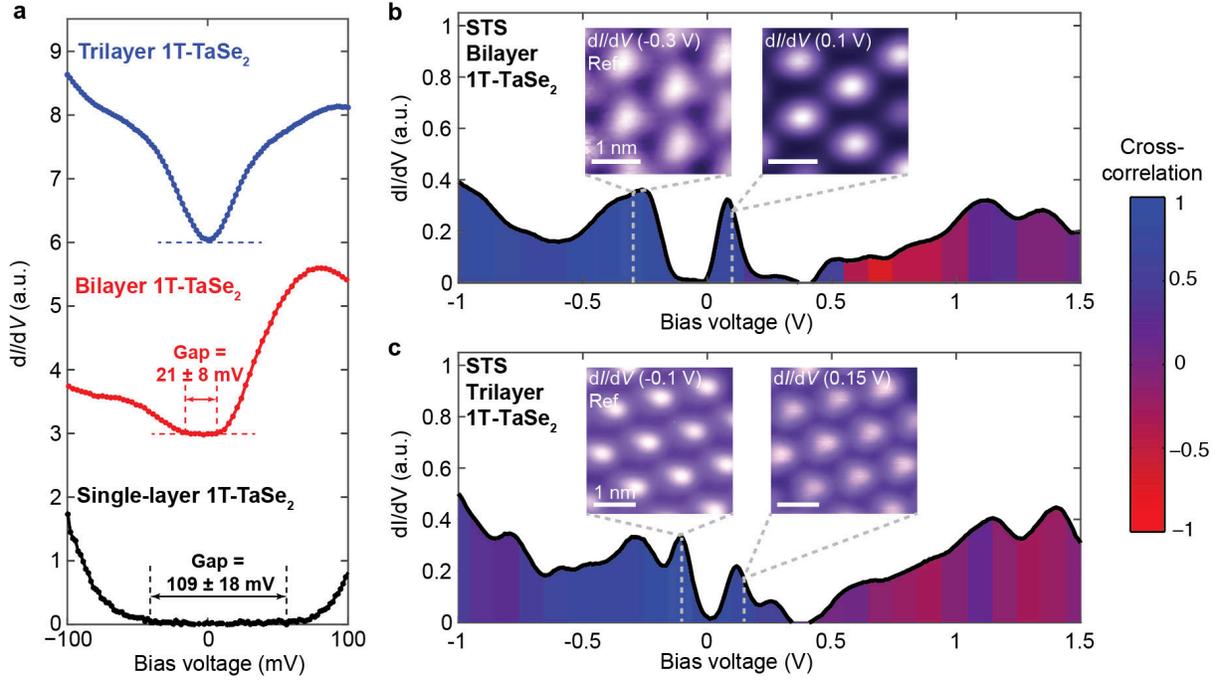

**Fig. 4. Energy gap reduction and quenching of exotic orbital texture in few-layer 1T-TaSe$_2$.**
**a**, STS d$I$/d$V$ spectra for single-layer, bilayer, and trilayer 1T-TaSe$_2$ show how interlayer coupling reduces the energy gap with an increasing number of layers. Spectra are shifted vertically for viewing (horizontal dashed lines mark d$I$/d$V$ = 0, $f$ = 401 Hz, $V_{RMS}$ = 2 mV). d$I$/d$V$ maps of the valence and conduction band LDOS as well as larger energy-scale d$I$/d$V$ spectra of **b,** bilayer, **c,** trilayer 1T-TaSe$_2$ ($f$ = 401 Hz, $V_{RMS}$ = 20 mV). Spatial cross-correlation values are shown color-coded with references taken near the LDOS maximum of the valence band for bilayer and trilayer 1T-TaSe$_2$. In contrast to SL1T-TaSe$_2$, the lowest conduction band for both bilayer or trilayer show no exotic orbital texture, thus resulting in positive cross-correlation values (blue), indicating that LDOS is concentrated on the interior Ta A- and B- atoms.

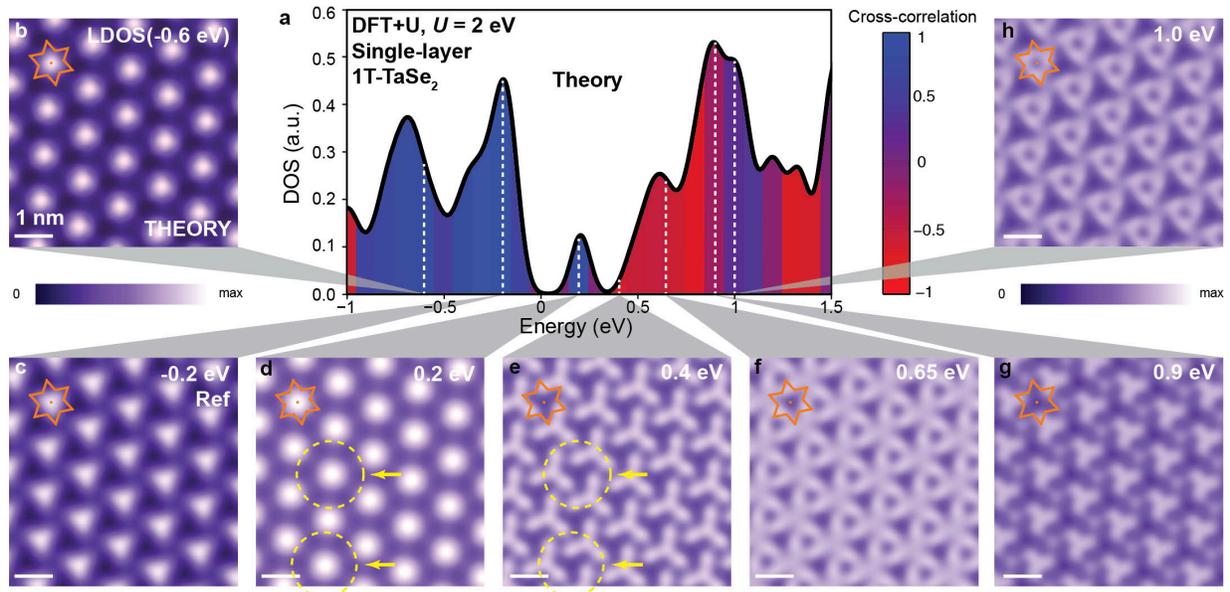

**Fig. 5. Theoretical orbital texture of single-layer 1T-TaSe$_2$ from DFT+U simulations. a**, Theoretical density of states of single-layer 1T-TaSe$_2$ from DFT+U simulations ($U$ = 2 eV). Color shows cross-correlation of LDOS maps at different energies with respect to the reference map in **c** (-0.2 eV). **b-h**, Theoretical LDOS maps of single-layer 1T-TaSe$_2$ from DFT+U simulations ($U$ = 2 eV). The same star-of-David supercell is outlined in each map (orange line). Yellow dashed circles in **d**, **e** highlight two star-of-David clusters which show very different theoretical conduction band orbital texture compared to experimental C$_1$ and C$_2$ features in Figs. 3d, e.

# Supplementary Information for

# Visualizing Exotic Orbital Texture in the Single-Layer Mott Insulator 1T-TaSe$_2$

Yi Chen[†], Wei Ruan[†], Meng Wu[†], Shujie Tang[†], Hyejin Ryu, Hsin-Zon Tsai, Ryan Lee, Salman Kahn, Franklin Liou, Caihong Jia, Oliver R. Albertini, Hongyu Xiong, Tao Jia, Zhi Liu, Jonathan A. Sobota, Amy Y. Liu, Joel E. Moore, Zhi-Xun Shen, Steven G. Louie, Sung-Kwan Mo, and Michael F. Crommie[*]

*† These authors contributed equally to this work.*

*\*e-mail: crommie@berkeley.edu*

**Table of Contents**

**Supplementary Notes**

**Part 1. Estimation of charge transfer between single-layer 1T-TaSe$_2$ and bilayer graphene**

When materials having different work functions are placed together, charge transfer can occur at their junction. An estimation of the charge transfer between 1T-TaSe$_2$ and graphene is necessary since the existence of a Mott insulating phase in 1T-TaSe$_2$ requires a band to be close to half filling. A charge transfer estimation was made possible by acquiring spatially-dependent STS spectra on the bilayer graphene (BLG) close to single-layer 1T-TaSe$_2$ islands (Supplementary Fig. 12). The Dirac point of bare BLG on SiC(0001) substrate is usually located at ~ -0.38 V due to substrate-induced n-doping[1,2]. If charge transfer between BLG and 1T-TaSe$_2$ exists, the n-doping of the graphene under the 1T-TaSe$_2$ island should be changed by this and its Dirac point shifted, while graphene far away from the island should be unaffected. Supplementary Fig. 12b shows spatially-dependent STS spectra of pristine BLG acquired along a line perpendicular to a single-layer 1T-TaSe$_2$ island. The low-energy phonon gap of graphene is clearly visible[2,3] in the STS spectrum acquired 5 nm away from the 1T-TaSe$_2$ island and its Dirac point is located at -0.37 V, in good agreement with previous reports[1,2]. As the tip moves closer to the 1T-TaSe$_2$ island, the Dirac point has a slight overall trend of shifting up and eventually reaches -0.34 V at the edge of the island. This energy shift corresponds to a change in charge density of ~3 × $10^{12}$ carriers/cm$^2$ in AB-stacked BLG[4]. As the flat band of single-layer 1T-TaSe$_2$ in DFT-GGA calculations hosts 1.5 × $10^{14}$ states/cm$^2$, this amount of charge transfer would dope it at most by ~2%. We thus conclude that the charge transfer from the bilayer graphene substrate to 1T-TaSe$_2$ is insignificant.

**Part 2. DFT+U simulations of single-layer 1T-TaSe$_2$**

In addition to the DFT+U results shown in the main text (with $U$ = 2 eV in the ferromagnetic (FM) state; see Figs. 2c and 5) that reproduce much of the experimentally-observed electronic structure, we also performed DFT+U calculations with different $U$ values. Supplementary Fig. 16 shows DFT+U band structures with $U$ values between 1 and 4 eV in the FM configuration (i.e., assuming that the electron at the center of each star-of-David is aligned ferromagnetically with the others across the material). With larger $U$ values, the electronic gap size around the Fermi level is found to increase, while higher-energy bands shift slightly. At $U$ = 2 eV, the overall DFT+U electronic structure and LDOS mapping show best agreement with experimental results. At $U$ > 5 eV, all Ta atoms become spin polarized and the electronic structure is dramatically modified.

The DFT+U results are not sensitive to the magnetic configurations adopted. This is shown by further DFT+U simulations of two representative antiferromagnetic (AFM) states on a triangular lattice, namely the striped AFM state (Supplementary Fig. 14) and the in-plane 120-degree AFM state (Supplementary Fig. 15). Both the striped AFM state ($U$ = 2 eV) and the 120-degree AFM state ($U$ = 2 eV) are found to open a gap at the Fermi level with a size similar to the FM state having the same $U$. In addition, both AFM states exhibit similar density of states and LDOS maps to those of the FM state and, in particular, no flower pattern emerges at the lowest conduction band peak at 0.2 eV (central Ta A-atom weights dominate instead).

**Part 3. Estimation of the potential landscape from occupied charge density in single-layer 1T-TaSe$_2$**

To experimentally estimate the electrostatic potential due to occupied charge density, we first probed the total occupied electron distribution by performing an STM constant-height current map at -3 V (Supplementary Fig. 18a). This reflects the energy-integrated LDOS distribution of the occupied states. To obtain the 2D charge distribution, further height-integration would be needed. This can be evaluated, however, by finding the spatial distribution of STM decay lengths from a grid of *I-Z* spectra. As shown in Supplementary Fig. 18c, the distribution of decay lengths exhibits high spatial homogeneity. This indicates that electronic wave-functions of the occupied states decay at roughly the same rate with tip height at different locations; thus, the charge density measured at a fixed height from the sample surface reflects the 2D charge density.

To determine the electronic potential due to occupied electronic states, we assume that the charge density distribution at every star-of-David forms a Gaussian profile (this fits the data reasonably well):

$$\rho_0(\boldsymbol{r}) = \frac{1}{2\pi\sigma^2} e^{-\frac{r^2}{2\sigma^2}}. \tag{1}$$

We then place these Gaussian-distributed charges into a periodic hexagonal lattice (i.e., the CDW superlattice), whose lattice vectors are $\{\mathbf{R} = m\mathbf{a}_1 + n\mathbf{a}_2 \mid m, n \in \mathbb{Z}\}$ where $\mathbf{a}_1, \mathbf{a}_2$ are the CDW unit vectors. We then fit this to our data to obtain the parameter $\sigma$ (the fitting result is shown in Supplementary Fig. 18b).

As in the Ewald method for computing Madelung energies, calculating the resulting electrostatic potential is simplified by working in Fourier space. The Fourier component of a single Gaussian-distributed charge $\rho_0(r)$ is

$$\rho_0(\mathbf{q}) = \int \rho_0(\mathbf{r})e^{-i\mathbf{q}\cdot\mathbf{r}}d^2\mathbf{r} = e^{-\frac{q^2\sigma^2}{2}}. \tag{2}$$

The resulting Fourier component of an entire periodic charge distribution is then

$$\rho(\mathbf{q}) = \int \rho(\mathbf{r})e^{-i\mathbf{q}\cdot\mathbf{r}}d^2\mathbf{r} = \int \sum_\mathbf{R} \rho_0(\mathbf{r}-\mathbf{R})e^{-i\mathbf{q}\cdot\mathbf{r}}d^2\mathbf{r} = \sum_\mathbf{R} \rho_0(\mathbf{q})e^{-i\mathbf{q}\cdot\mathbf{R}}. \tag{3}$$

Next, we compute the electrostatic potential induced by this periodic charge distribution. The Coulomb potential $V_0(\mathbf{r}) = e^2/(4\pi\epsilon r)$ has a Fourier component of

$$V_0(\mathbf{q}) = \int V_0(\mathbf{r})e^{-i\mathbf{q}\cdot\mathbf{r}}d^2\mathbf{r} = \frac{e^2}{4\pi\epsilon}\frac{2\pi}{q} = \frac{e^2}{2\epsilon q}. \tag{4}$$

We can write the total electrostatic potential as a spatial convolution

$$V(\mathbf{r}) = \int \rho(\mathbf{r}')V_0(\mathbf{r}-\mathbf{r}')d^2\mathbf{r}', \tag{5}$$

thus allowing us to obtain the Fourier components of the potential:

$$V(\mathbf{q}) = \rho(\mathbf{q})V_0(\mathbf{q}) = \int V(\mathbf{r})e^{-i\mathbf{q}\cdot\mathbf{r}}d^2\mathbf{r}. \tag{6}$$

The resulting real-space potential is

$$V(\mathbf{r}) = \frac{1}{(2\pi)^2}\int V(\mathbf{q})e^{i\mathbf{q}\cdot\mathbf{r}}d^2\mathbf{q}. \tag{7}$$

Since the potential has the same translational symmetry as the crystal ($V(\mathbf{r}) = V(\mathbf{r}+\mathbf{R})$), it can be simplified as a discrete Fourier sum

$$V(\mathbf{r}) = \sum_\mathbf{G} V_\mathbf{G} e^{i\mathbf{G}\cdot\mathbf{r}}. \tag{8}$$

Here $\{\mathbf{G}\}$ are the reciprocal lattice vectors and the discrete Fourier components are

$$V_{\mathbf{G}} = \frac{1}{V_{\text{U.C.}}} \int_{\text{U.C.}} V(\mathbf{r}) e^{-i\mathbf{G}\cdot\mathbf{r}} d^2\mathbf{r}, \tag{9}$$

where $V_{\text{U.C.}} = \sqrt{3}|a_{1,2}|^2/2 = \sqrt{3}a^2/2$ is the volume of the unit cell. The continuous and discrete Fourier components are linked by

$$V(\mathbf{q}) = \int V(\mathbf{r}) e^{-i\mathbf{q}\cdot\mathbf{r}} d^2\mathbf{r} = \sum_{\mathbf{R}} \int_{\text{U.C.}} V(\mathbf{r}+\mathbf{R}) e^{-i\mathbf{q}\cdot(\mathbf{r}+\mathbf{R})} d^2\mathbf{r} = \sum_{\mathbf{R}} V_{\text{U.C.}} V_{\mathbf{q}} e^{-i\mathbf{q}\cdot\mathbf{R}}, \tag{10}$$

and we know that

$$V(\mathbf{q}) = \rho(\mathbf{q}) V_0(\mathbf{q}) = \sum_{\mathbf{R}} \rho_0(\mathbf{q}) e^{-i\mathbf{q}\cdot\mathbf{R}} V_0(\mathbf{q}). \tag{11}$$

Using the relation that $\mathbf{G}\cdot\mathbf{R}$ is a multiple of $2\pi$,

$$V(\mathbf{G}) = \sum_{\mathbf{R}} V_{\text{U.C.}} V_{\mathbf{G}} = \sum_{\mathbf{R}} \rho_0(\mathbf{G}) V_0(\mathbf{G}), \tag{12}$$

allows us to obtain

$$V_{\mathbf{G}} = \frac{1}{V_{\text{U.C.}}} \rho_0(\mathbf{G}) V_0(\mathbf{G}) = \frac{e^2}{\sqrt{3}\epsilon a^2 G} e^{-\frac{G^2\sigma^2}{2}}. \tag{13}$$

The real-space potential landscape can thus be expressed as

$$V(\mathbf{r}) = \sum_{\mathbf{G}} V_{\mathbf{G}} e^{i\mathbf{G}\cdot\mathbf{r}} = \sum_{\mathbf{G}} \frac{e^2}{\sqrt{3}\epsilon a^2 G} e^{-\frac{G^2\sigma^2}{2}} e^{i\mathbf{G}\cdot\mathbf{r}}, \tag{14}$$

which is the equation we use for obtaining Supplementary Fig. 18d (we take a G cutoff of 10 reciprocal lattice vectors). The resulting periodic electrostatic potential that we estimate for single-layer 1T-TaSe$_2$ displays minima at the flower-petal locations between the stars-of-David (compare Supplementary Figs. 18d and 18e).

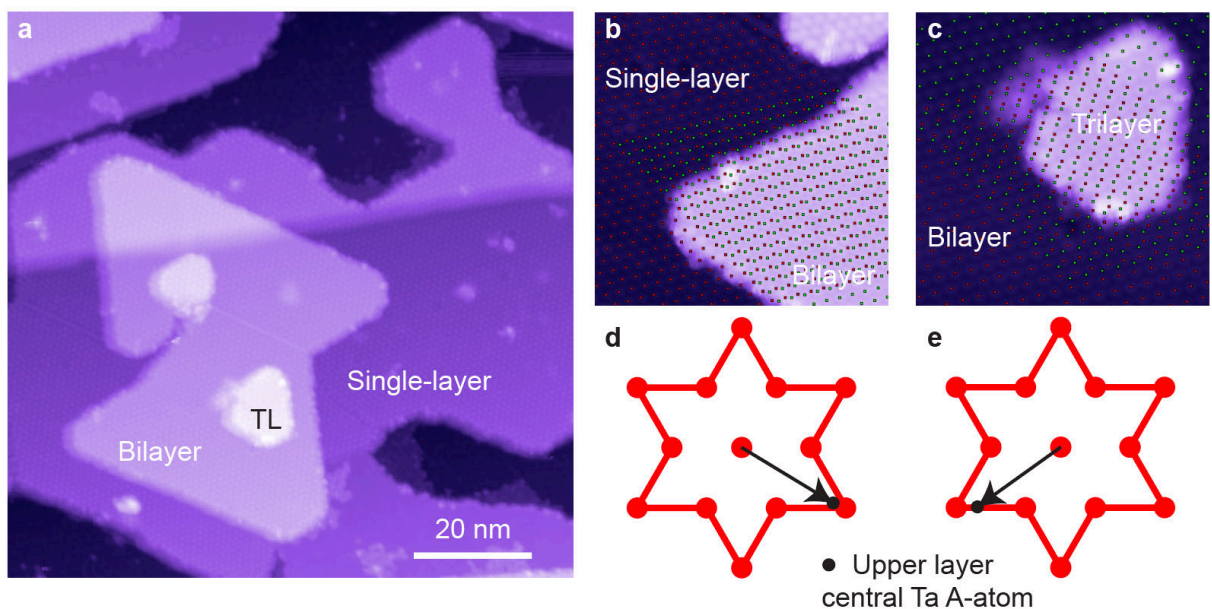

**Supplementary Figure 1. Shifted (i.e., triclinic) stacking observed in 1T-TaSe$_2$ thin films. a**, Overall STM topograph showing an island consisted of single-layer, bilayer, and trilayer 1T-TaSe$_2$. **b-c**, Zoom-in of topographs used for finding the stacking order. Red (green) dots mark the centers of star-of-David CDW cells in the lower (upper) layer. **d**, Schematic of observed shifted stacking between single-layer and bilayer 1T-TaSe$_2$. Here the central Ta A-atom in the upper layer (black circle) sits near a peripheral Ta-C atom in the lower layer (red star-of-David). **e**, Schematic of observed shifted stacking between bilayer and trilayer 1T-TaSe$_2$ plotted in the same fashion as **d**.

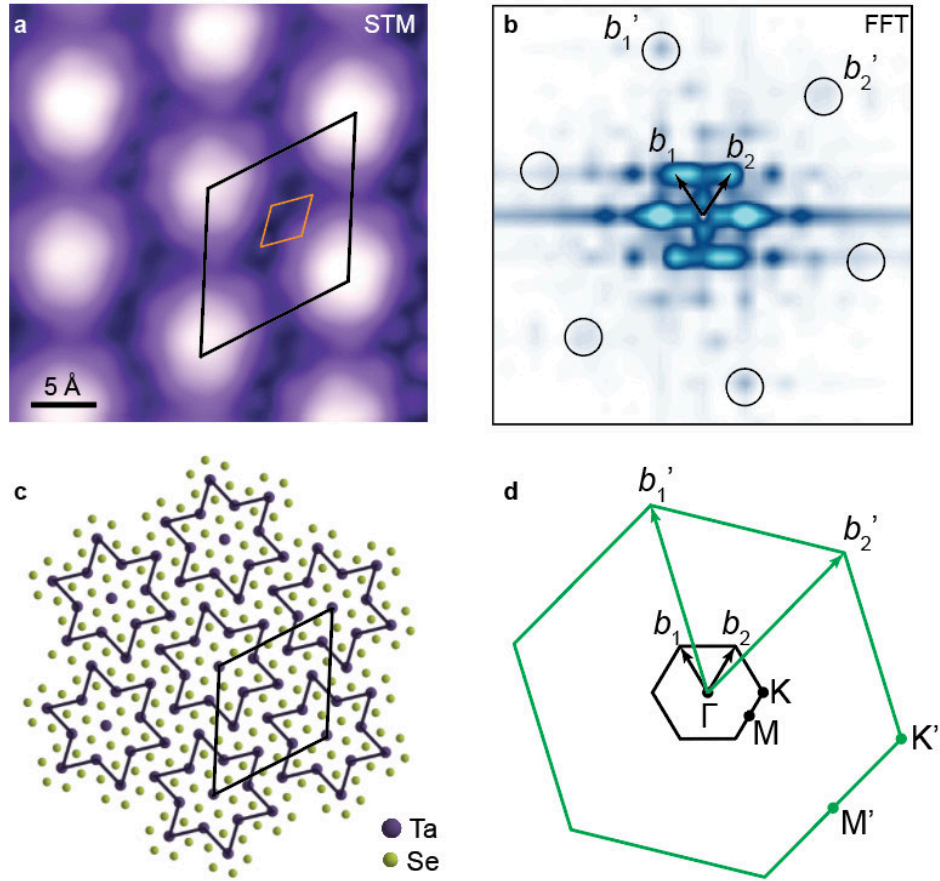

**Supplementary Figure 2. Fourier analysis of measured atomic and CDW lattices in single-layer 1T-TaSe$_2$. a**, STM image of single-layer 1T-TaSe$_2$. **b**, Fourier transform of the image in **a**. Peaks of both the CDW superlattice (marked by $b_1$ and $b_2$) and the atomic lattice (i.e., undistorted lattice) ($b_1'$ and $b_2'$) are visible. The atomic lattice constant is 3.41 $\pm$ 0.16 Å and the CDW superlattice lattice constant is 12.40 $\pm$ 1.63 Å. The CDW superlattice is rotated by an angle of ~14 degree from the atomic lattice. The uncertainty here is determined by the half-width at half maximum of the Fourier peaks. **c**, Sketch of the atomic and CDW lattices. **d**, 2D Brillouin zones and reciprocal lattice vectors of the atomic and CDW lattices.

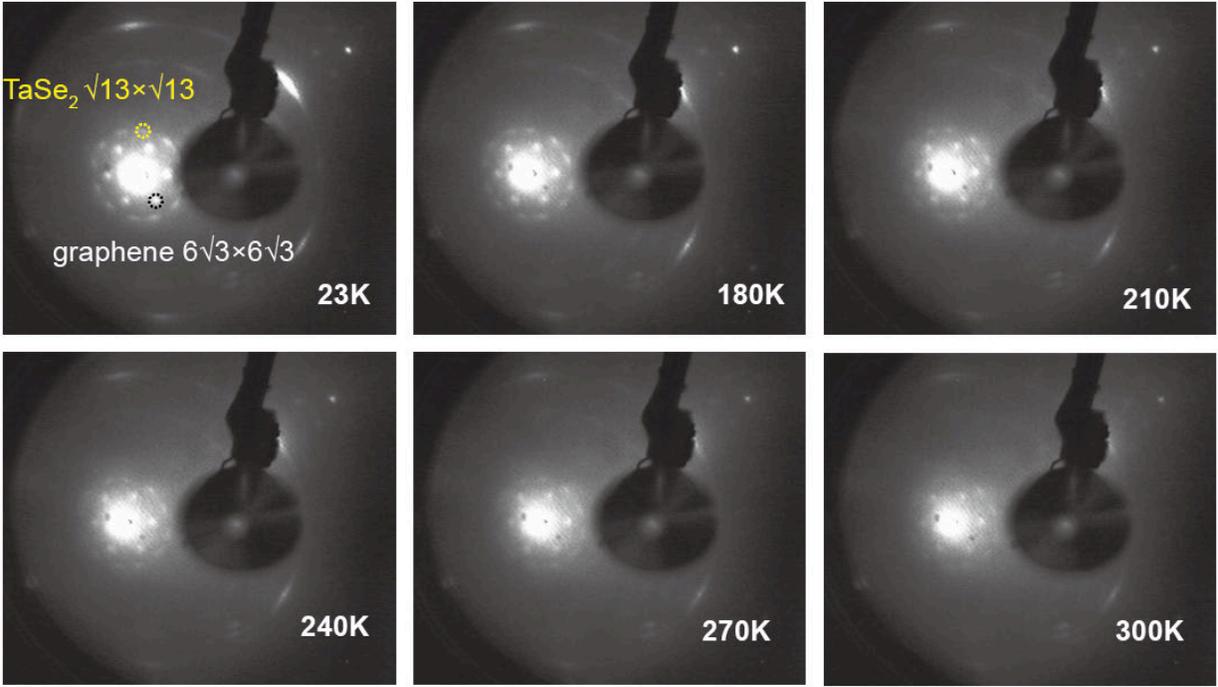

**Supplementary Figure 3. Temperature-dependent LEED patterns for single-layer 1T-TaSe$_2$.** The $\sqrt{13} \times \sqrt{13}$ CDW superlattice of single-layer 1T-TaSe$_2$ is observed in LEED patterns at all measured temperatures.

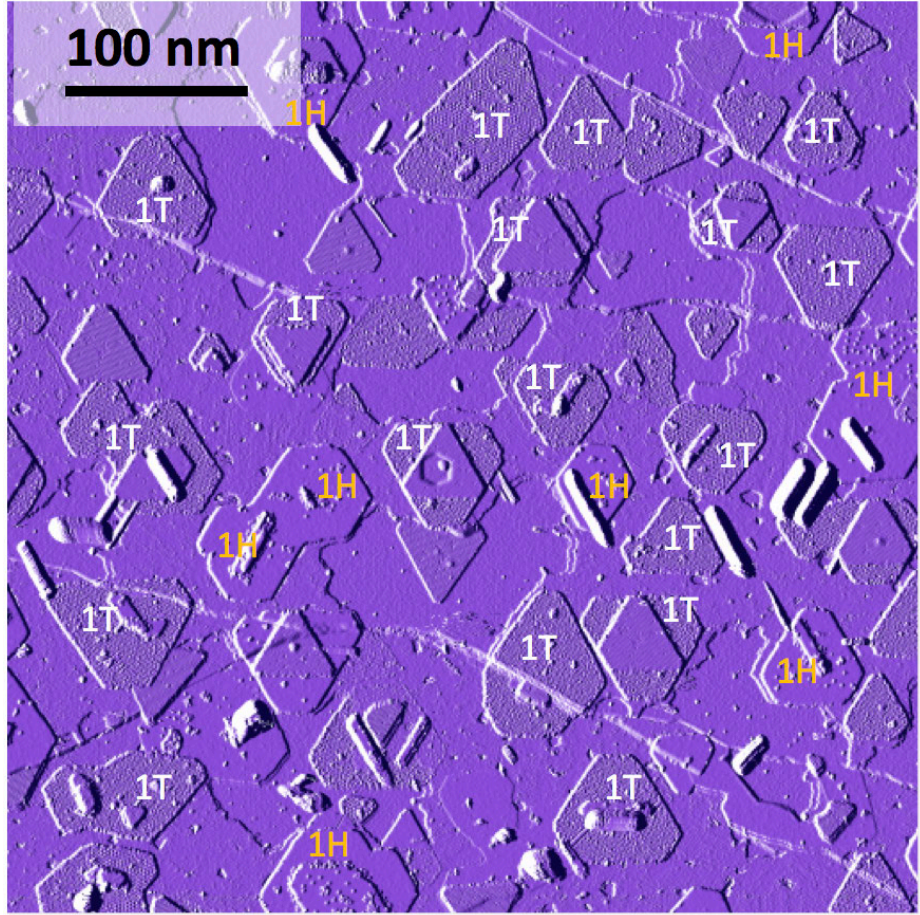

**Supplementary Figure 4. Large-scale STM topograph showing coexisting 1T- and 1H-TaSe$_2$ islands.** Large-scale STM image identifying 1T and 1H islands ($V_b$ = 0.5 V, $I_t$ = 10 pA) (image intensity here is proportional to the derivative of the height with respect to x in order to enhance contrast between islands for this large image).

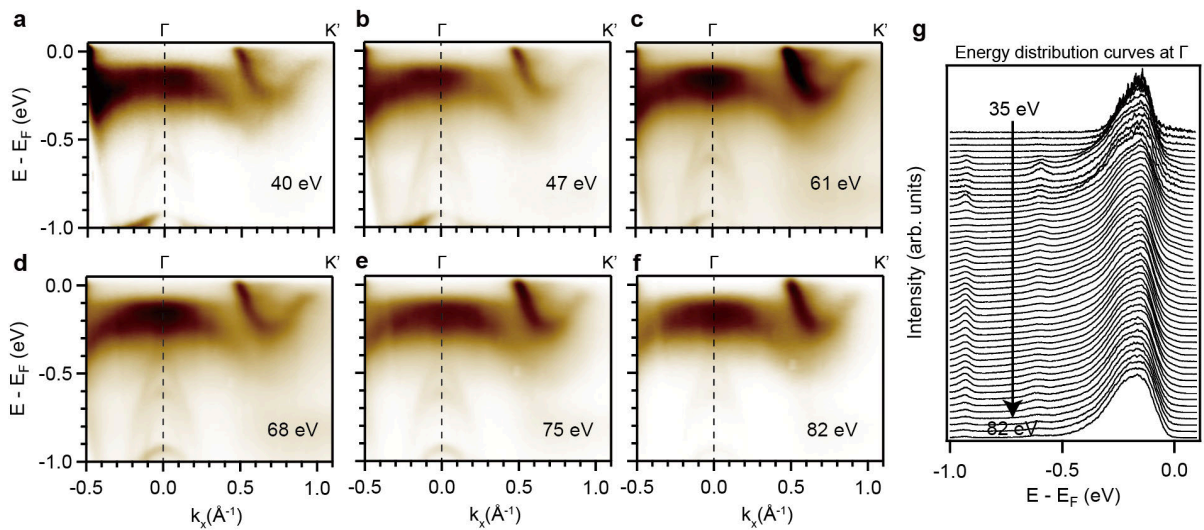

**Supplementary Figure 5. Photon-energy dependence of ARPES spectra for single-layer 1T-TaSe$_2$. a-f,** ARPES spectra along the Γ-K' direction taken with *p*-polarized light at different photon energies. **g**, Energy distribution curves (EDCs) at the Γ point acquired with photon energies ranging from 35 eV to 82 eV. The ARPES spectra and EDC show no obvious photon energy dependence.

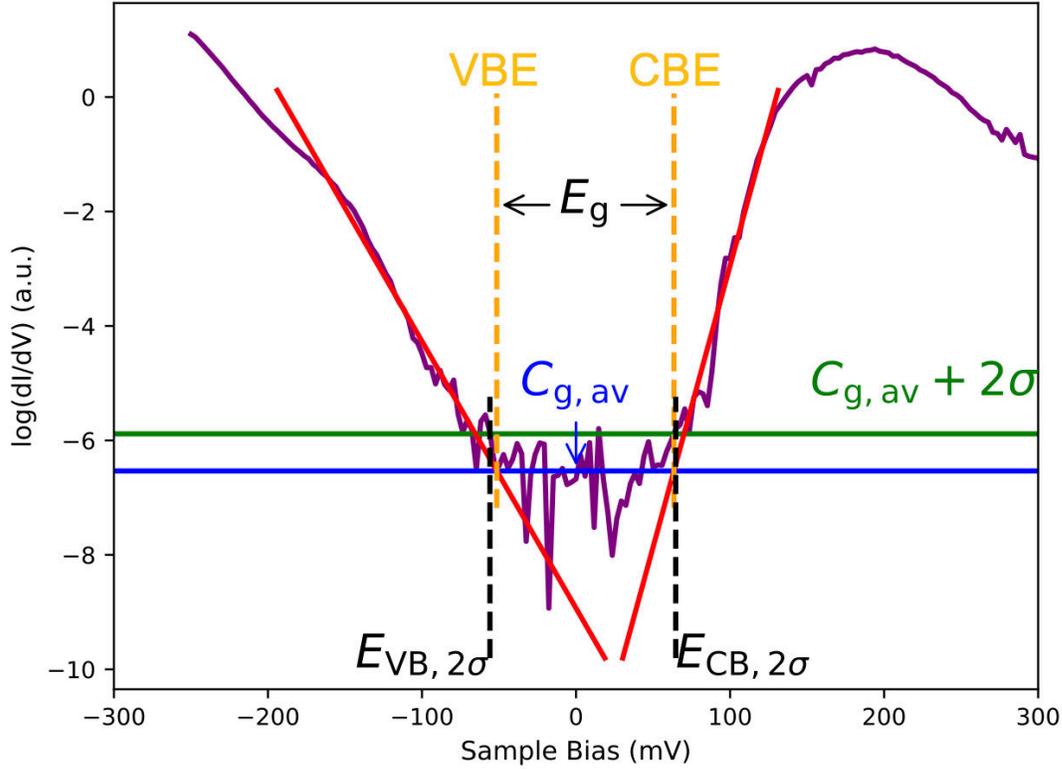

**Supplementary Figure 6. Determination of the STS gap size for single-layer 1T-TaSe$_2$.** The STS gap size was determined from the logarithm of 24 individual STS curves using a similar algorithm as in ref. [5]. One such curve is shown in purple. We first selected an energy window of [-50, 50] mV, within which the mean of the signal was calculated ($C_{g,av}$, blue line) together with its standard deviation ($\sigma$). The green line marks two standard deviations above $C_{g,av}$, whose intersections with the STS curve are denoted by $E_{VB,2\sigma}$ and $E_{CB,2\sigma}$. Beyond $E_{VB,2\sigma}$ and $E_{CB,2\sigma}$, linear fits (red lines) were made over energy windows chosen so that the R$^2$ values of the linear fits are greater than 0.95. The gap edges (VBE and CBE) are defined as the energies at which the linear fits intersect $C_{g,av}$, whose difference yields the gap size.

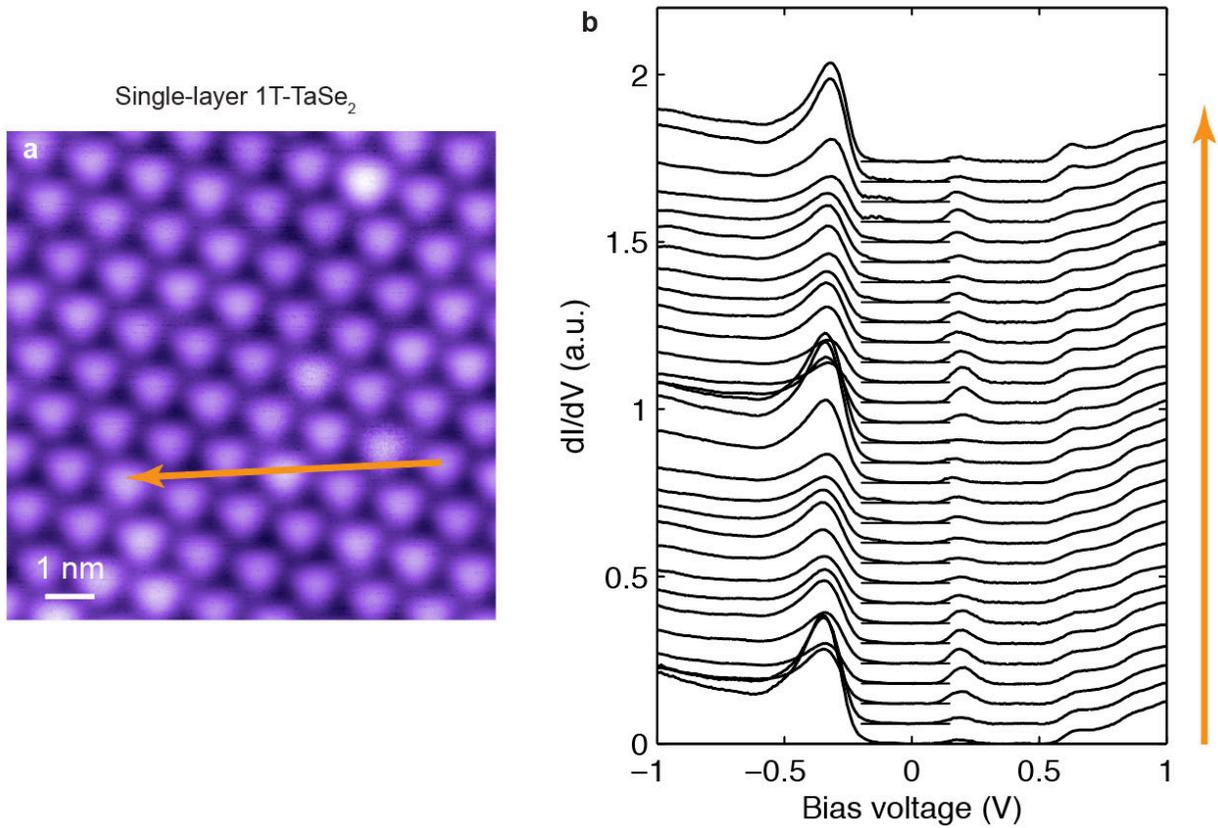

**Supplementary Figure 7. Spatially-dependent STS spectra for single-layer 1T-TaSe$_2$ showing spatial homogeneity of the gapped electronic structure. a**, STM topograph of a pristine region of single-layer 1T-TaSe$_2$ ($V_b$ = -0.4 V, $I_t$ = 10 pA). The orange arrow shows where STS line spectra were acquired. **b**, STS spectra acquired along orange arrow show spatially-uniform gapped electronic structure across several CDW superstructures. Some variation can be seen in the amplitudes of the peaks ($f$ = 401 Hz, $I_t$ = 30 pA, $V_{RMS}$ = 20 mV). The position of zero in each spectrum is marked by a horizontal black line.

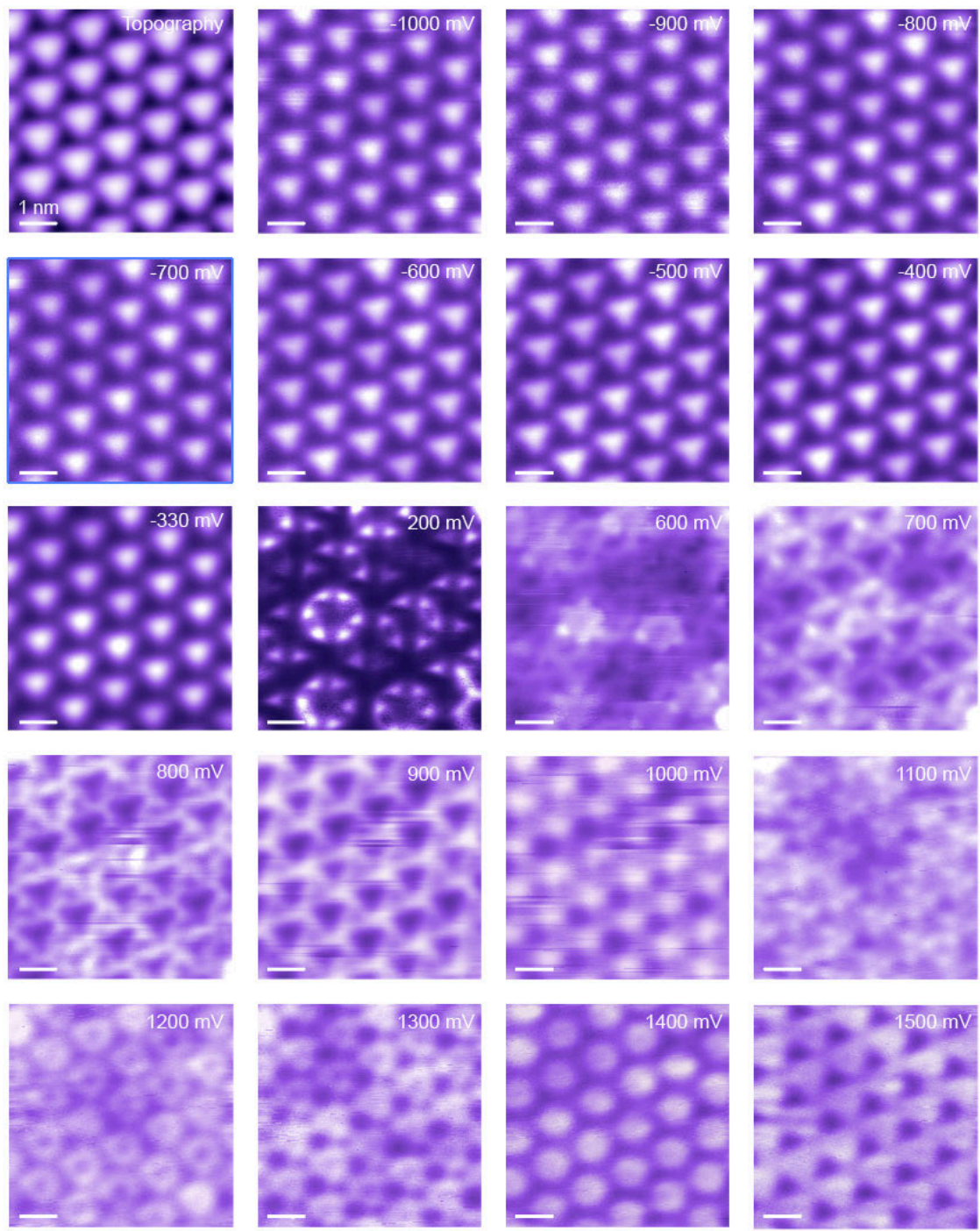

**Supplementary Figure 8. Constant-height d$I$/d$V$ conductance maps of single-layer 1T-TaSe$_2$ at different biases** ($f$ = 401 Hz, $V_{RMS}$ = 20 mV).

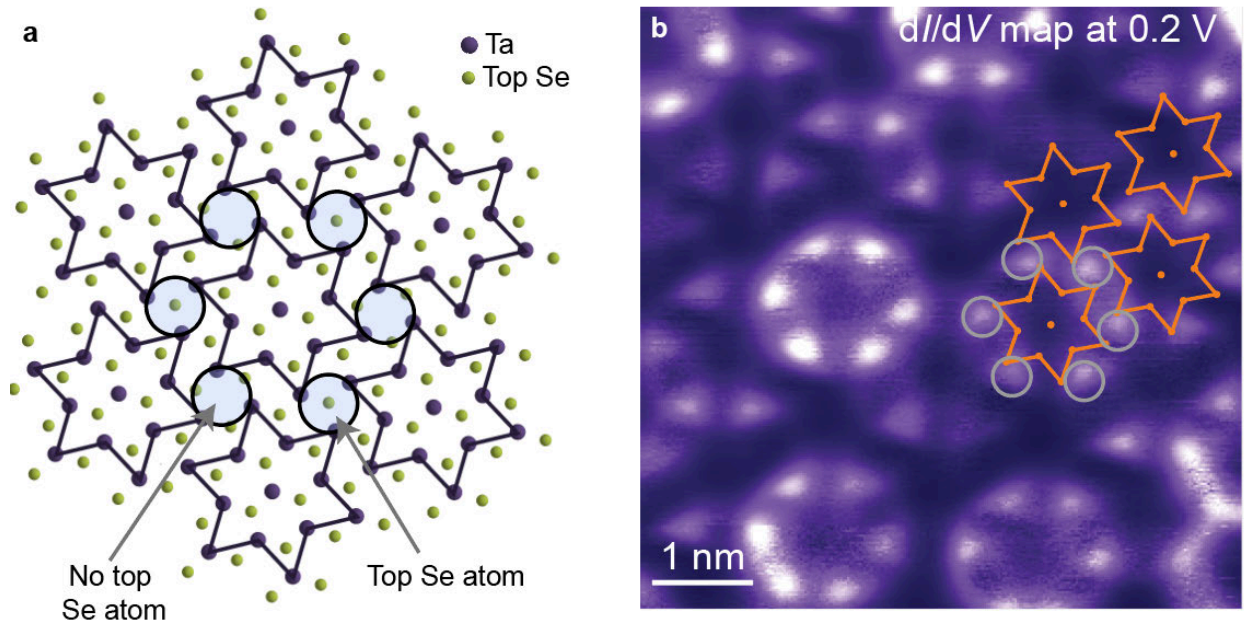

**Supplementary Figure 9. Correspondence between atomic sites and petals in the flower-like LDOS map at $C_1$ (0.2 V) for single-layer 1T-TaSe$_2$. a**, Sketch shows Ta atoms and Se atoms in the *top* Se plane of 1T-TaSe$_2$. The top Se plane exhibits three-fold symmetry and so cannot account for the six-fold symmetric petals of the conduction band LDOS at $C_1$ (light blue shading). Ta C-atoms, however, do have the proper symmetry and are thus mostly likely responsible for the flower-like LDOS pattern. **b**, d$I$/d$V$ map of $C_1$ conduction band state of single-layer 1T-TaSe$_2$ at $V$ = 0.2 V with outline of star-of-David supercells added and locations of "flower petals" marked by circles.

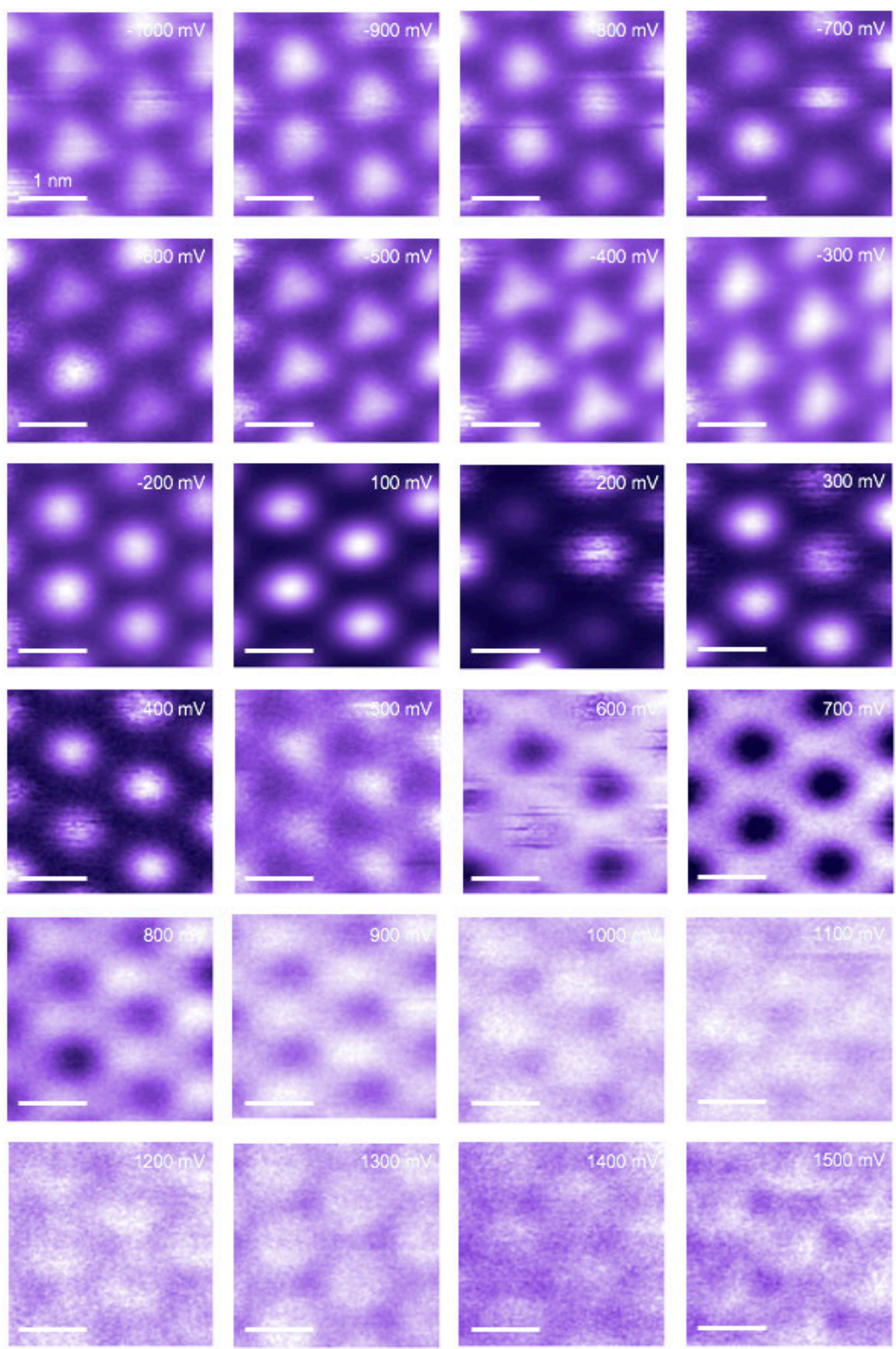

**Supplementary Figure 10. Constant-height d*I*/d*V* conductance maps of bilayer 1T-TaSe$_2$ at different biases** ($f$ = 401 Hz, $V_{RMS}$ = 20~30 mV).

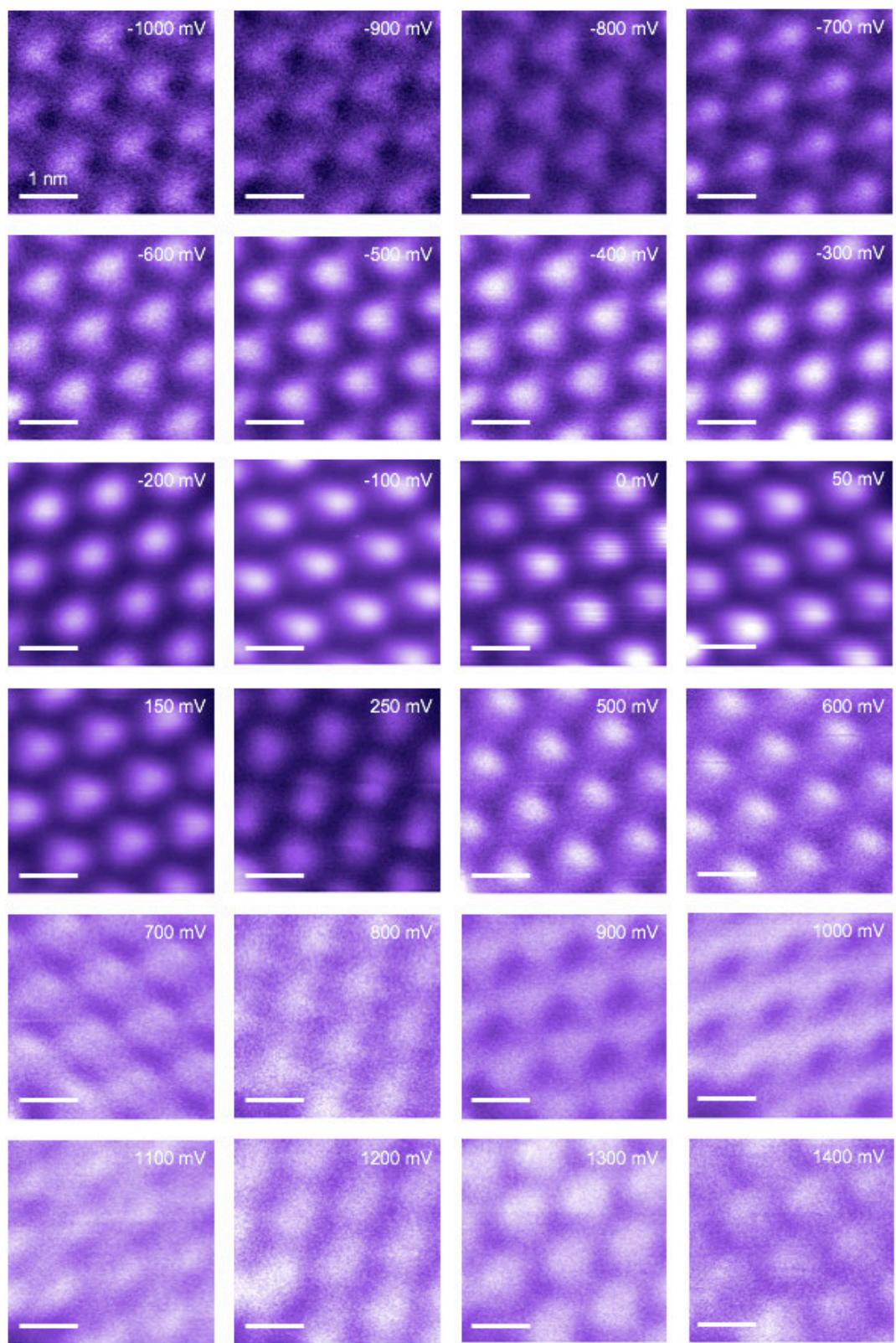

**Supplementary Figure 11. Constant-height d$I$/d$V$ conductance maps of trilayer 1T-TaSe$_2$ at different biases** ($f$ = 401 Hz, $V_{RMS}$ = 20 mV).

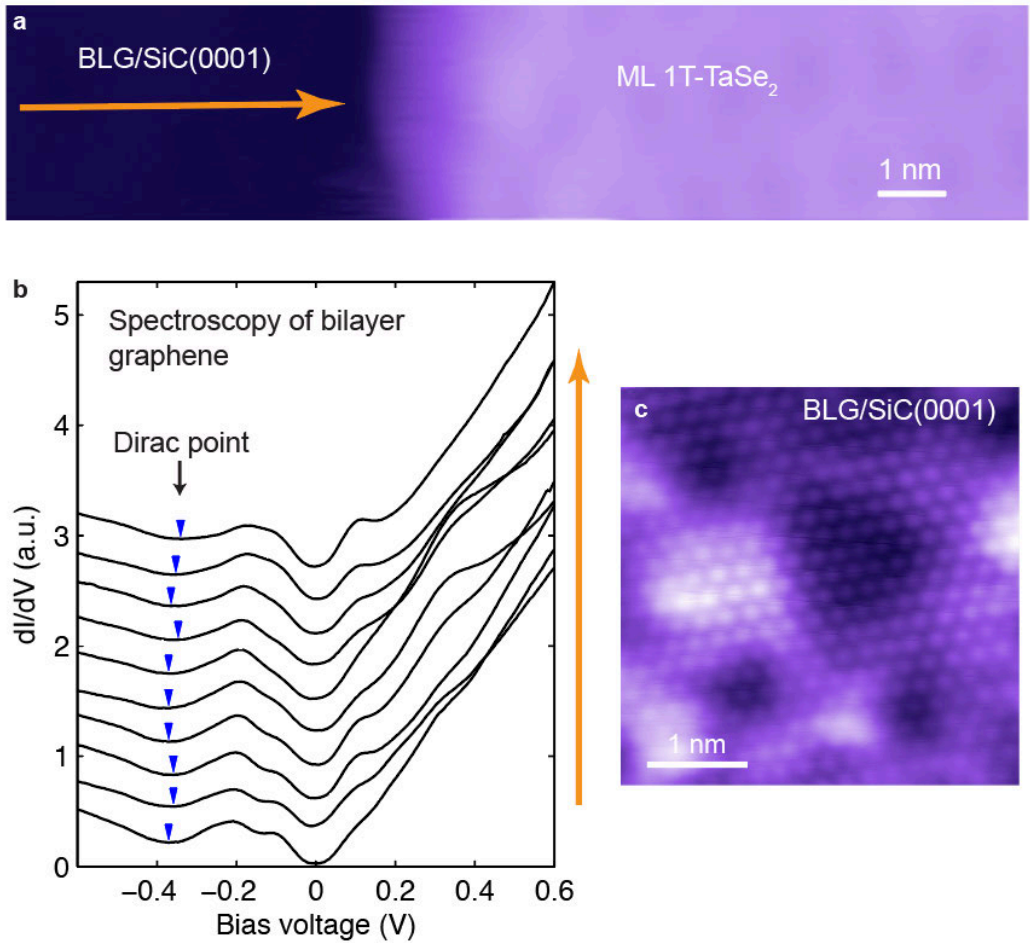

**Supplementary Figure 12. Spatially-dependent STS spectra of bilayer graphene close to a single-layer 1T-TaSe$_2$ island. a**, STM topograph of investigated area. Orange arrow shows where spectra were obtained ($V_b$ = -0.6 V, $I_t$ = 10 pA). **b**, Spatially-dependent STS spectra of bilayer graphene obtained along orange arrow shown in **a** ($f$ = 401 Hz, $I_t$ = 50 pA, $V_{RMS}$ = 30 mV). Dirac points are marked by blue triangles. **c**, Close-up STM image of the bilayer graphene substrate shows its honeycomb structure ($V_b$ = 0.05 V, $I_t$ = 100 pA).

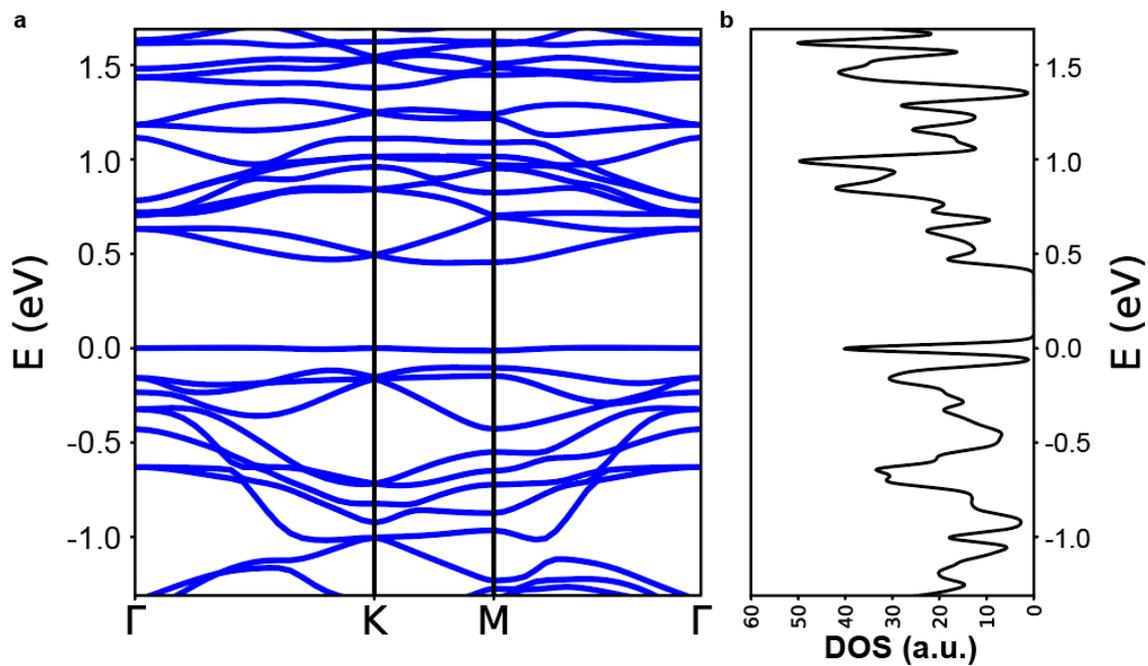

**Supplementary Figure 13. DFT-GGA band structure of single-layer 1T-TaSe$_2$. a**, Band structure of single-layer 1T-TaSe$_2$ at the DFT-GGA level shows a half-filled narrow band at the Fermi level (0 eV). **b,** Density of states in the same calculation.

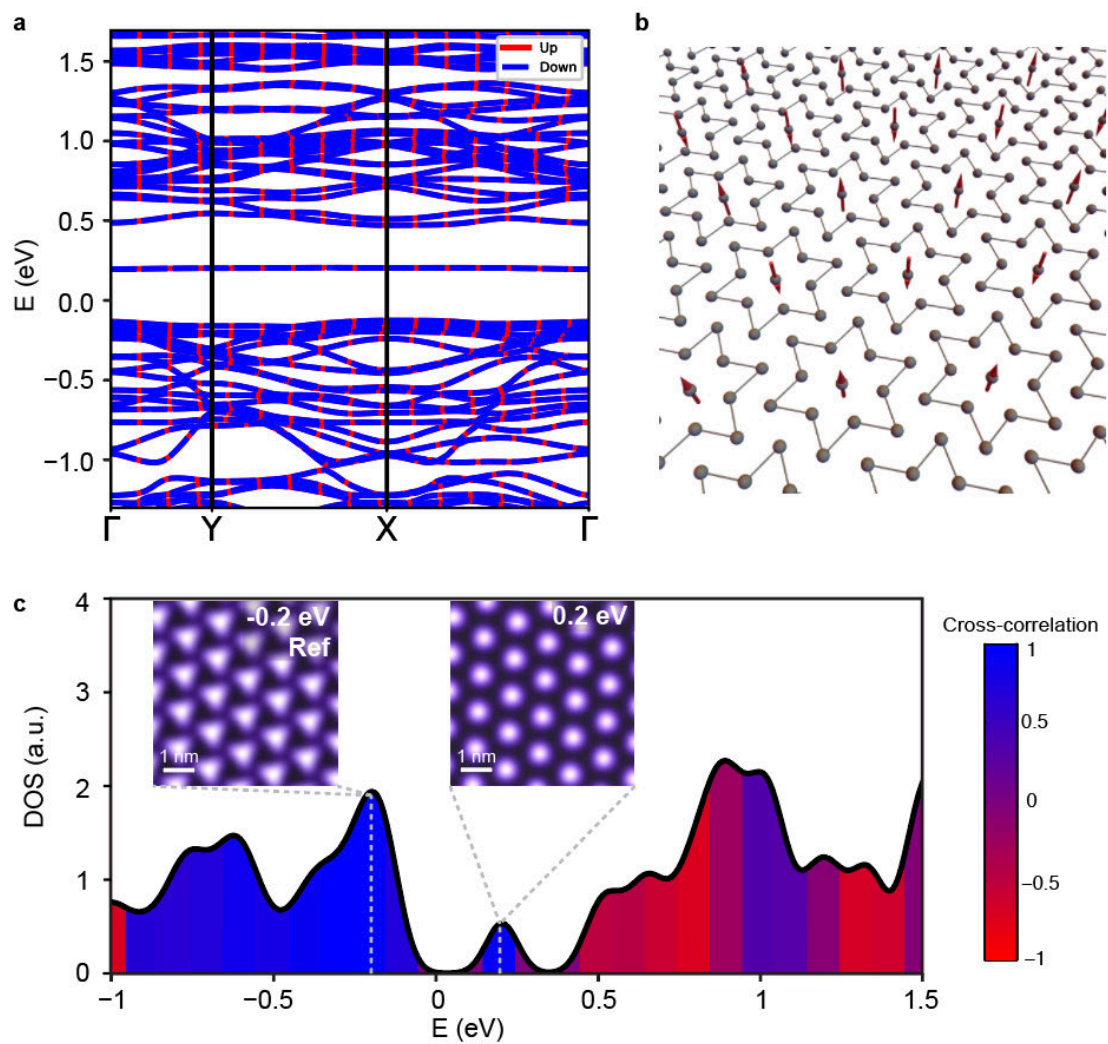

**Supplementary Figure 14. DFT+U band structure of single-layer 1T-TaSe$_2$ in the striped antiferromagnetic state with $U$ = 2 eV. a**, DFT+U band structure calculated for the striped AFM Brillouin zone. **b,** Schematic of spins in the striped AFM state. **c,** Density of states from the same calculation with color-coded cross-correlation with the reference LDOS map at -0.2 eV. Inset: LDOS maps at -0.2 eV and 0.2 eV show the central Ta atom weights dominating.

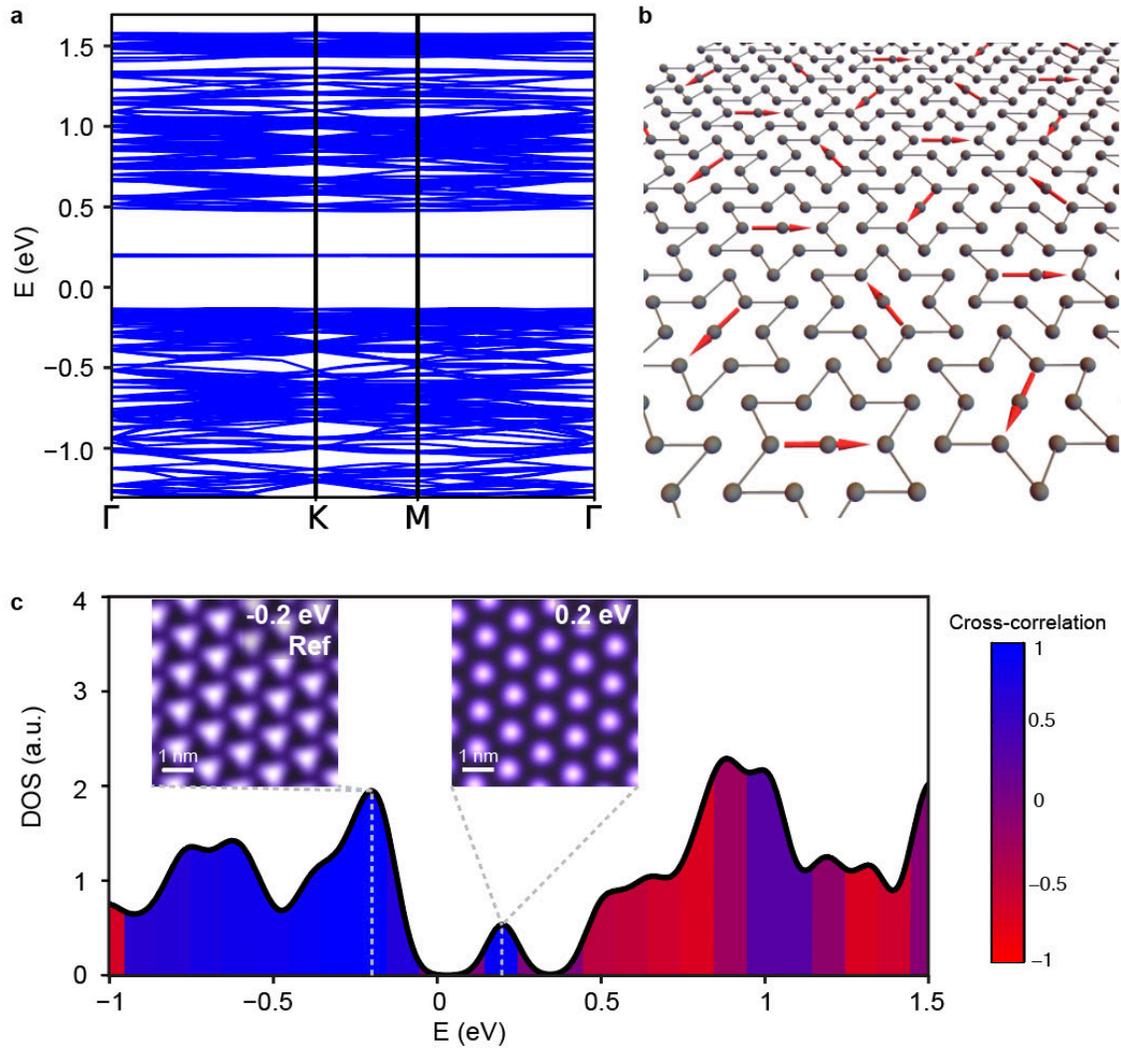

**Supplementary Figure 15. DFT+U band structure of single-layer 1T-TaSe$_2$ in the in-plane 120-degree antiferromagnetic state with $U = 2$ eV. a**, DFT+U band structure in the 120-degree AFM Brillouin zone. **b,** Schematic of spins in the 120-degree AFM state. **c,** Density of states from the same calculation with color-coded cross-correlation with the reference LDOS map at -0.2 eV. Inset: LDOS maps at -0.2 eV and 0.2 eV show the central Ta atom weights dominating.

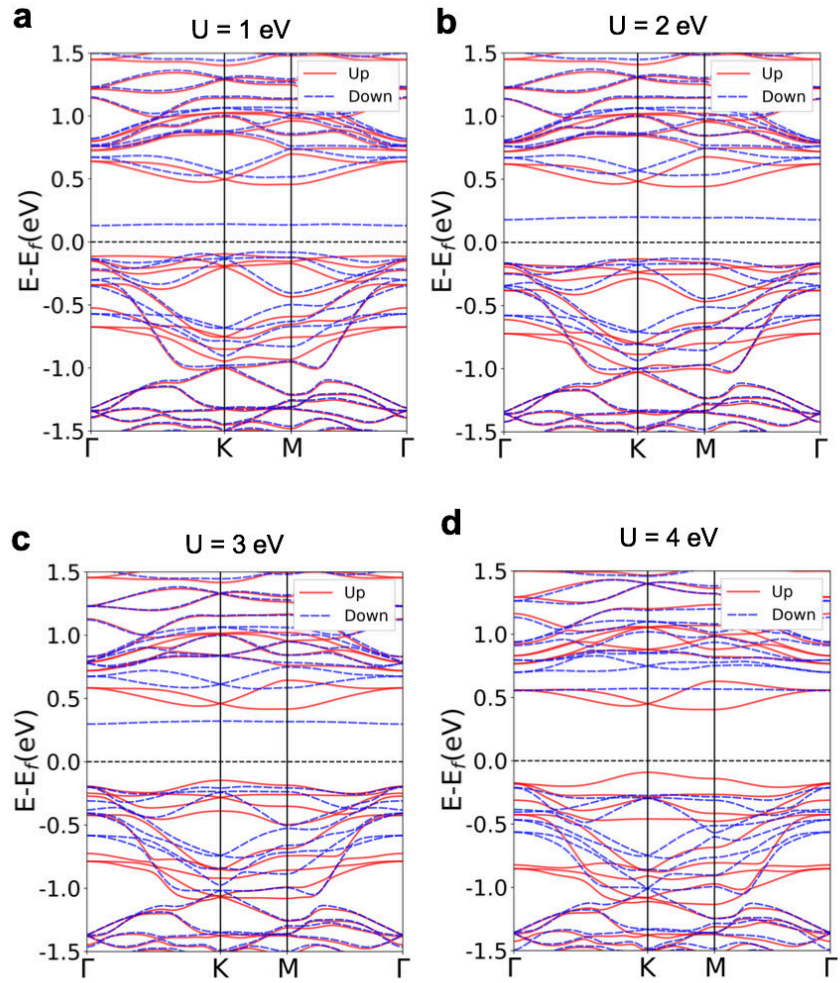

**Supplementary Figure 16. DFT+U band structure of single-layer 1T-TaSe$_2$ with different *U* values in the ferromagnetic state.** DFT+U band structures corresponding to **a**, *U* = 1 eV, **b**, *U* = 2 eV, **c**, *U* = 3 eV, and **d**, *U* = 4 eV. The *U* = 2 eV band structure shows best agreement with the experimentally-observed electronic structure of single-layer 1T-TaSe$_2$ and is used in the main text.

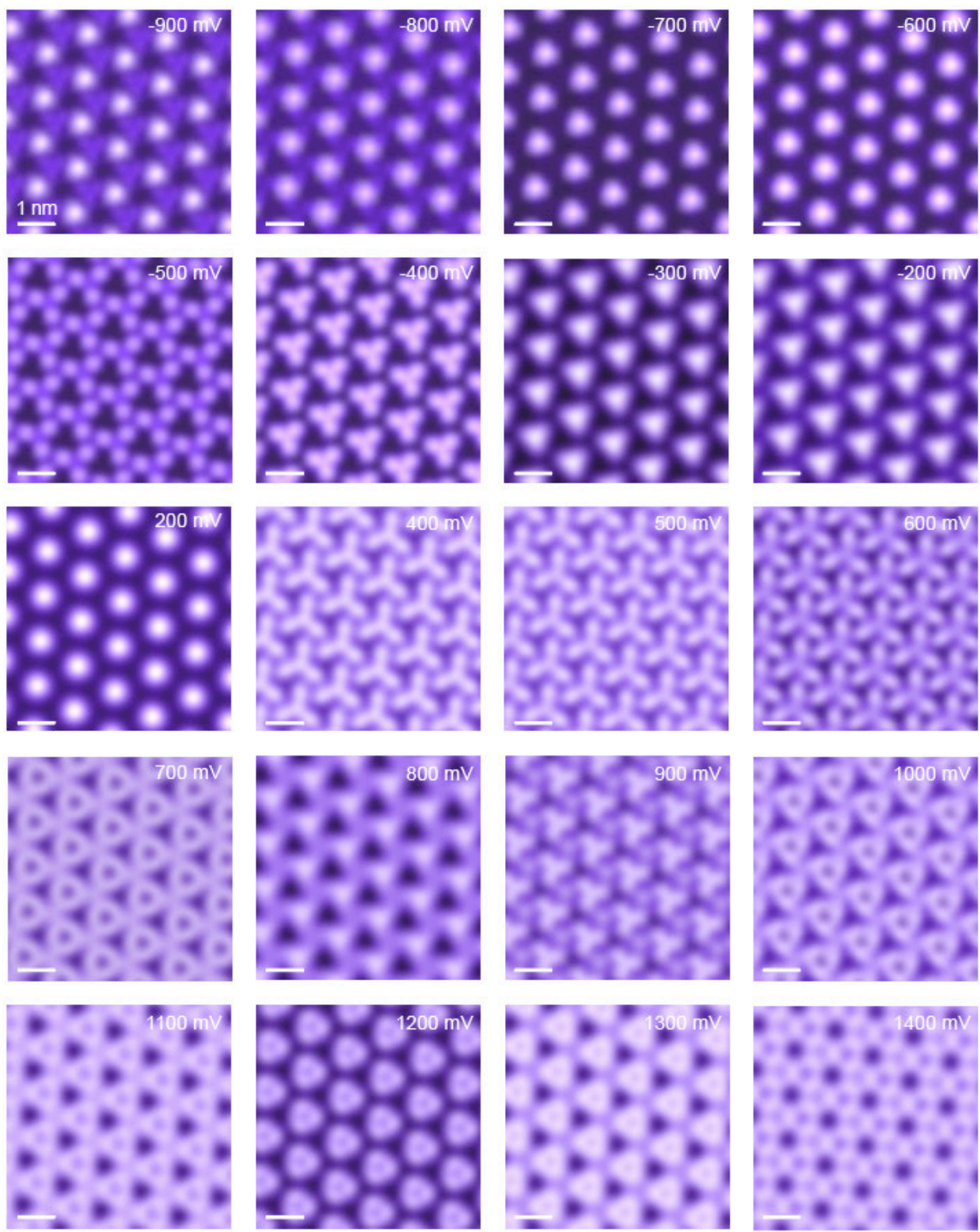

**Supplementary Figure 17. Theoretical LDOS maps from DFT+U calculation ($U = 2$ eV, ferromagnetic state) for single-layer 1T-TaSe$_2$.**

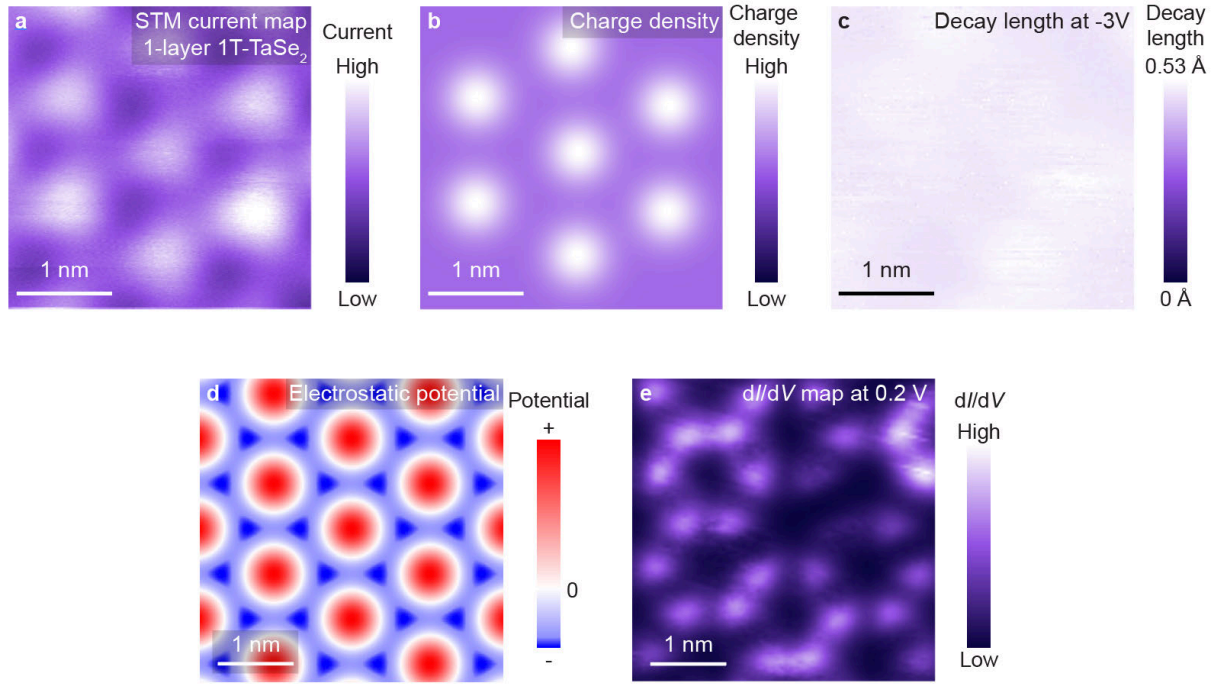

**Supplementary Figure 18. Estimation of charge density and electrostatic potential from STM measurements of single-layer 1T-TaSe$_2$. a**, STM constant-height current map at $V = -3$ V. **b**, Gaussian fit of the charge density from **a**. **c**, Spatially-resolved decay length at $V = -3$ V measured by fitting *I-Z* curves acquired over this area shows that decay length has no significant spatial dependence. **d**, Electrostatic potential calculated from charge density according to Equation 14. **e**, Flower-like LDOS pattern observed at 0.2 V ($C_1$) in the STM d*I*/d*V* map of single-layer 1T-TaSe$_2$. The positions of the bright "flower petals" in **e** match well with the electrostatic potential minima (blue regions) in **d**.